\documentclass[showpacs,preprintnumbers,amssymb,twocolum,aps,reprint,superscriptaddress,showkeys,amsmath,floatfix]{revtex4-2}

\usepackage{graphicx}
\usepackage{dcolumn}
\usepackage[dvipsnames]{xcolor}
\usepackage{bm}
\usepackage{amsmath}
\usepackage{amssymb}
\usepackage{epsfig}
\usepackage{amsfonts}
\usepackage{lineno,hyperref}
\usepackage{array}
\usepackage{booktabs}
\usepackage{float}
\usepackage{microtype}
\usepackage{multirow}
\usepackage{adjustbox}
\usepackage[english]{babel}
\usepackage{epstopdf}
\usepackage{blindtext}
\usepackage{booktabs}
\usepackage{subcaption}
\usepackage[a4paper, total={6.5in, 10in}]{geometry}
\usepackage{appendix}

\def \l{\lambda}
\def \L{\Lambda}

\def \o{\omega}
\def \O{\Omega}

\def \D{\Delta}

\def \t{\theta}

\def \be{\begin{equation}}
\def \ee{\end{equation}}
\def \ben{\begin{eqnarray}}
\def \een{\end{eqnarray}}

\def \n{\nonumber}

\def \La{\mathcal{L}}

\def \t{\Tilde}

\newcommand{\SHOES}{SH0ES}

\newcommand{\PantheonShoes}{Pantheon+\SHOES}

\begin{document}

\title{Observational Insights on DBI  K-essence Models Using Machine Learning and Bayesian Analysis}

\author{Samit Ganguly}
\email{samitgphy07@gmail.com}
\affiliation{Department of Physics, University of Calcutta, 92, A.P.C. Road, Kolkata-700009, India}
\affiliation{Department of Physics, Haldia Government College, Haldia, Purba Medinipur 721657, India}

\author{Arijit Panda}
\email{arijitpanda260195@gmail.com}
\affiliation{Department of Physics, K. J. Somaiya College of Science and Commerce, Vidya Vihar, Mumbai 400077, India}

\author{Eduardo Guendelman}
\email{guendel@bgu.ac.il}
\affiliation{Department of Physics, Ben-Gurion University of the Negev, Beer-Sheva, Israel}
\affiliation{Frankfurt Institute for Advanced Studies (FIAS), Ruth-Moufang-Strasse 1, 60438 Frankfurt am Main, Germany.}
\affiliation{Bahamas Advanced Study Institute and Conferences, 4A Ocean Heights, Hill View Circle, Stella Maris, Long Island, The Bahamas.}

\author{Debashis Gangopadhyay}
\email{debashis.g@snuniv.ac.in}
\affiliation{Department of Physics, School of Natural Sciences, Sister Nivedita University, DG 1/2, Action Area 1, Newtown, Kolkata 700156, India.}

\author{Abhijit Bhattacharyya}
\email{abhattacharyyacu@gmail.com}
\affiliation{Department of Physics, University of Calcutta, 92, A.P.C. Road, Kolkata-700009, India}

\author{Goutam Manna$^a$}
\email{goutammanna.pkc@gmail.com \\$^a$Corresponding author}
\affiliation{Department of Physics, Prabhat Kumar College, Contai, Purba Medinipur 721404, India} 
\affiliation{Institute of Astronomy, Space and Earth Science, Kolkata 700054, India}

\date{\today}

\begin{abstract} 
We perform a late-time cosmological study, we compare the performance of two Dirac–Born–Infeld (DBI) type k-essence scalar field extensions of the $\Lambda$CDM model to the standard framework and a $w$CDM scenario using the Chevallier–Polarski–Linder (CPL) equation of state parametrization. We solve background dynamics numerically as functions of redshift and incorporate them into a Bayesian inference pipeline accelerated by machine learning. We use a Flax-based surrogate emulator to replace repeated direct integrations of the ODE system, reducing computational cost. A hybrid scheme that combines Stochastic Variational Inference (SVI) with No-U-Turn Hamiltonian Monte Carlo constrains cosmological parameters using the \PantheonShoes{} Type Ia supernova sample, DESI BAO (DR2) data, and cosmic chronometer $H(z)$ measurements without CMB-based priors. In both DBI k-essence formulations, present-day dark energy equations of state are consistent with cosmic acceleration, indicating a $\Lambda$CDM-like regime with a modest redshift dependence. The $w$CDM model is marginally favored by conventional model selection measures such as $\chi^2$, AIC, BIC, and DIC, which are based on goodness of fit and penalized. However, Bayesian predictive measures like WAIC and PSIS–LOO show no significant differences between $\Lambda$CDM, $w$CDM, and DBI k-essence scenarios. All have similar model weights and out-of-sample predictive performance for the datasets. Thus, DBI k-essence models mimic the success of the classic $\Lambda$CDM paradigm while allowing controlled, redshift-dependent deviations from a strict cosmological constant that are consistent with present late-time observations.
\end{abstract}

\keywords{%
K-essence geometry, Dark energy, Dark matter, Late time acceleration, Observational data analysis}

\pacs{04.20.-q, 04.50.Kd, 98.80.-k, 98.80.Es} 

\maketitle


\section{Introduction}
The discovery that the expansion of the universe is not slowing down, as once expected, but instead accelerating, stands as one of the most profound revelations in modern cosmology \cite{Perlmutter, Riess, Tegmark, Spergel, Scranton, Mainini}. This acceleration, first inferred from observations of distant Type Ia supernovae and later confirmed through measurements of the cosmic microwave background (CMB) and large-scale structure, challenges our understanding of gravity and the cosmic behavior of the universe. Within the framework of general relativity, elucidating this phenomenon necessitates the introduction of a new energy component, dark energy, which comprises about $70\%$ of the universe's total energy density and exerts a negative pressure sufficient to produce cosmic acceleration. 

The cosmological constant, albeit the most straightforward and well examined option for dark energy, has significant theoretical difficulties. The standard cosmological model, the $\Lambda$CDM model, faces two main issues: fine-tuning and the cosmic coincidence problem. The cosmological constant $\Lambda$ is a significant factor in the observed energy density, causing a 120 order difference in predictions and observations ~\cite{Weinberg, Peebles, Padmanabhan2, Panda2, Panda1}. This significant disparity, known as the \emph{cosmological constant problem}, has prompted the creation of many alternative theories. 

Although the fine tuning and cosmic coincidence problems pose enduring conceptual challenges in cosmology, they are not typically regarded as direct inconsistencies of the $\Lambda$CDM model. As a result, many researchers view them as theoretical concerns that may require reinterpretation rather than a fundamental modification of the prevailing paradigm \cite{Linder, Victor}. In contrast, the Hubble tension refers to the discrepancy between early and late universe measurements of the Hubble constant $H_0$. The early universe estimate from CMB data, such as that from the Planck collaboration \cite{Planck5}, is $H_0 \approx 67.4 \pm 0.5\, \mathrm{km/s/Mpc}$, whereas late universe direct local measurements, particularly from the SH0ES team \cite{Riess, Riess1}, yield $H_0 \approx 73.0 \pm 1.0\, \mathrm{km/s/Mpc}$. This discrepancy has become a statistically significant observational issue. Its resolution likely necessitates new physics beyond the standard cosmological model. Proposed possibilities often include either altering the matter 
composition of the universe or revisiting the basic principles of gravity. Compounding these tensions, recent BAO measurements from the Dark Energy Spectroscopic Instrument Data Release 2 (DESI DR2) have reported $\sim 2$--$3\sigma$ evidence for evolving dark energy within the CPL parametrization~\cite{Desi, Hussain2}, a result further examined and corroborated by several independent analyses~\cite{Chaudhary, Capozziello, Capozziello1}. These findings collectively motivate the exploration of dynamical dark energy models beyond the cosmological constant, and provide additional observational impetus for the DBI k-essence framework investigated in this work.

In this context, alternative theories of gravity have emerged as compelling avenues for understanding late time acceleration. These theories aim to go beyond Einstein's general relativity by introducing new degrees of freedom or modifying the behavior of gravity on cosmological scales, often without the need for a finely tuned cosmological constant. Among them, scalar field theories have received considerable attention \cite{Ratra, Ferreira, Brax1, Brax2, Brax3}, particularly those involving noncanonical kinetic terms that enrich the dynamics and offer greater flexibility in modeling the universe's expansion history.

A significant category of these models is K-essence \cite{Picon1, Picon2, Babichev1, dg1, dg2, dg3, Vikman, Sawicki, Chimento, Visser, Scherrer, Padmanabhan, Mukhanov1, Picon3, Chiba, Mukohyama, Das, gm2, gm3, gm4, gm5, gm6, gm7, gm8, Ganguly, Panda, Panda1, Panda2, Panda3, García, Csillag},  wherein the Lagrangian exhibits a nonlinear dependence on both the scalar field and its kinetic term. The K-essence theory is a particular subclass of generalized scalar tensor theories, situated within the context of Horndeski gravity~\cite{Horndeski, Deffayet, Joyce}. It incorporates scalar fields with non canonical kinetic parameters and offers a dynamic framework to elucidate cosmic acceleration while avoiding higher order instabilities. K-essence, first suggested in relation to inflation, has since been modified to elucidate dark energy and the acceleration of the universe in its latter stages. These models are especially appealing due to their capacity to demonstrate dynamical attractor solutions, allowing for a natural onset of acceleration without extreme sensitivity to initial conditions. Furthermore, K-essence permits a variable equation of state ($w$), perhaps aligning more closely with data than a constant $w$ model. K-essence theory may also be used to examine unified dark energy and dust dark matter \cite{Guendelman1} in the realms of inflation and dark energy \cite{Guendelman2} within an alternate framework. The formation of K-essence dynamics without fine tuning has been proven within the Two Measures Field Theory (TMT) framework, which offers a natural basis for scalar field cosmology \cite{Guendelman}. The detailed motivations, significance, applications, and various forms of the Lagrangian adapted to different cosmological scenarios within non canonical theories such as K-essence can be found in~\cite{Panda1, Panda2}.

The impetus for observational insights into k-essence theory inside FLRW backgrounds arises from the need to elucidate the universe's late time acceleration beyond the cosmological constant. Examining the effects on the FLRW metric facilitates evaluating the theory's coherence with empirical facts like supernovae, baryon acoustic oscillations, and the cosmic microwave background. As precision cosmology advances, investigating k-essence offers an essential understanding of cosmic acceleration and evaluates its feasibility as a model for the evolving universe.

The Dirac-Born-Infeld (DBI) type non canonical Lagrangian was originally inspired by attempts to resolve the problem of electron self energy in classical electrodynamics \cite{Born}. Its characteristic square root form naturally limits field gradients, preventing divergences and introducing nonlinear dynamics suitable for cosmology. In modern cosmological contexts, DBI  type k-essence has been applied in various scenarios, most notably in early universe inflation and late time cosmic acceleration. In inflation, it enables slow roll dynamics with distinctive signatures like large CMB non Gaussianities, while in dark energy, its low sound speed affects structure formation compared to canonical models \cite{Mukhanov1}. Additionally, DBI  type k-essence can also emerge from D brane dynamics in wrapped string geometries \cite{Dvali, Tong, Tong1, Roy, Guendelman3, Nozari}.

From a physical standpoint, DBI  type k-essence models are particularly compelling because they maintain a subluminal and stable sound speed throughout the evolution of the universe. The sound speed in such models is given by $c_s^2= \frac{\partial P}{\partial X}/\frac{\partial \rho}{\partial X}$, which, for DBI  models, remains well behaved and avoids instabilities like ghost modes or gradient instabilities that often afflict other dark energy scenarios with noncanonical kinetic terms \cite{Vikman, Mukhanov1}. This makes DBI-inspired k-essence models both theoretically appealing and observationally viable, offering a rich phenomenology without sacrificing stability. Our focus will be on understanding how these models can drive cosmic acceleration while preserving desirable features such as dynamical attractors, theoretical consistency, and a well behaved sound speed throughout cosmic history. 

In this study, we apply Bayesian sampling through Stochastic Variational Inference (SVI) combined with the No-U-Turn Sampler (NUTS) \cite{Hoffman}, together with contemporary Machine Learning (ML) techniques \cite{Ntampaka}, to examine cosmological observations using two distinct DBI-type k-essence scalar field models. ML has recently become a key tool in cosmology because of its capacity to capture complex structures and deliver accurate inferences from large data volumes, with supervised methods, such as decision trees, support vector machines, and neural networks, commonly employed for both classification and regression tasks. Complementarily, state-of-the-art sampling schemes like NUTS offer a rigorous probabilistic framework for parameter inference. NUTS, an adaptive extension of Hamiltonian Monte Carlo (HMC) \cite{Betancourt}, efficiently traverses high-dimensional posterior spaces without manual step-size or trajectory tuning, making it particularly advantageous when robust uncertainty quantification is required \cite{Gelman}. In this work, we do not consider CMB measurements and thereby restricting our analysis to late time cosmological datasets only.

This paper is organized as follows: In Section II, we introduce the two DBI type k-essence Lagrangians for scalar field models, derive their background dynamics, and formulate the essential evolution equations. Section III presents the observational framework and evolution equations in terms of redshift distance ($z$), including luminosity distance, Hubble parameter, and energy density evolution, adapted for both models. In Section IV, we detail the methodology, combining Bayesian inference via \texttt{Stochastic Variational Inference} using NumPyro with ML based surrogate modeling to fit the models to \PantheonShoes{}, Hubble and  BAO data  and reconstruct the EoS parameters using symbolic regression. Section V provides an extensive statistical comparison of the proposed models based on multiple criteria, including the chi-squared statistic ($\chi^2$), reduced chi-squared ($\chi^2_{\nu}$), Akaike Information Criterion (AIC), Bayesian Information Criterion (BIC), and Deviance Information Criterion (DIC). Since AIC and BIC depend only on the maximum likelihood, they offer limited information about global predictive performance. To obtain a more robust and Bayesian consistent model evaluation, we additionally use the Watanabe Akaike Information Criterion (WAIC) and Pareto-Smoothed Importance Sampling Leave-One-Out cross-validation (PSIS-LOO). These diagnostics perform a pointwise analysis of the log likelihood for each data from the Baryon Acoustic Oscillation (BAO) and cosmic chronometer samples, while treating the SN compilation as a single data point with its full covariance matrix, thereby enabling us to do a more comprehensive statistical comparison. This section also investigates the redshift evolution of the deceleration parameter and tests the physical viability of the models. Section VI then summarizes the main findings and discusses their consequences for k-essence cosmology and the dynamics of the dark sector.

\section{DBI  type K-essence lagrangian and derivation of essential dynamics}

K-essence theories are scalar field models characterized by a Lagrangian that depends nonlinearly on the kinetic term $X = \partial_{\mu}\phi\partial^{\mu}\phi$, where $\phi$ is the k-essence scalar field. Unlike canonical scalar field theories where the kinetic term appears linearly, k-essence allows for a richer dynamical structure, making it suitable for modeling both early and late time cosmic acceleration \cite{Scherrer, Padmanabhan, Mukhanov1, Vikman, Martin, Hussain1}. Among the many forms of k-essence, one particularly well-motivated example is the DBI-type k-essence model. In this work, we explore two distinct classes of k-essence scalar field models within the framework of the DBI-type Lagrangian. It is worth clarifying that the term \emph{k-essence theory}, as used throughout this paper, is standard in the literature~\cite{Picon1, Picon2, Mukhanov1, Vikman}. Fundamentally, it refers to a class of models where the kinetic energy of the scalar field typically dominates over the potential energy to drive the cosmological dynamics. Because of this, k-essence does not denote a single field configuration, but rather a well-defined \emph{class} of models that share the same underlying structure - specifically, a nonlinear kinetic Lagrangian. Each model within this framework admits its own action principle, equations of motion, stability criteria, attractor solutions, and observational signatures. In this sense, it constitutes a genuine theoretical framework, much like quintessence or Galileon gravity, and we therefore refer to it here as a ``theory''.\\

{\bf Model-1:} A general DBI type action is given by \cite{Vikman,Martin} :
\ben
S_{DBI}^{I} &&= -\int d^4x \sqrt{-g} ~ \Big[f(\phi)\sqrt{1- \frac{2X}{f(\phi)}}\n \\&&- f(\phi) + V(\phi)\Big]
\label{1}
\een
Here, $f(\phi)$ denotes the tension/wrap factor, which we refer to as the tension scalar, and $V
(\phi)$ is the potential.  Note that here we assume the scalar field to be homogeneous, i.e, $\phi(r,t) \equiv \phi(t)$ having {\it mass dimension [viz. Appendix \ref{F}]}, since k-essence geometry permits us to choose the field to be of the homogeneous type \cite{gm3, gm4}.  Now the variation of the above action with respect to the scalar field $\phi$ and the gravitational metric in the usual FLRW background with zero curvature leads to the scalar field equation of motion (EoM) and the energy momentum tensor $(T_{\nu(\phi)}^{\mu})$. We can express them respectively as:

\ben
&&\Ddot{\phi}- \frac{3 f'(\phi)}{2 f(\phi)}\dot{\phi}^2+ f'(\phi)+\frac{3H}{\gamma^2}\dot{\phi}\n \\&&+ \frac{1}{\gamma^3}[V'(\phi)-f'(\phi)]=0 
\label{2}
\een
and
\ben
T_{\nu(\phi)}^{\mu} = (\rho_{\phi} + P_{\phi}) u^{\mu}u_{\nu}- P_{\phi} \delta^{\mu}_{\nu} ,
\label{3}
\een
with 
\ben
&& u_{\mu}=\frac{\partial_{\mu} \phi}{\sqrt{2X}} \quad ;\quad  u_{\mu}u^{\mu}=1. 
\label{4}
\een
Here, in Eq. (\ref{2}) $'dash'$ denotes the derivative with respect to the scalar field $\phi$ and $'dot'$ denotes the derivative with respect to coordinate time, $t$ and (\ref{3}) $\rho_{\phi}$ and $P_{\phi}$ denote the energy density and pressure of the scalar field sector, respectively. $H~(=\frac{\dot{a}}{a})$ is the Hubble parameter, $a\equiv a(t)$ is the scale factor. We can write the total energy density and pressure as \cite{Martin}:

\ben
&&\rho= \rho_{\phi} = (\gamma - 1) f(\phi) +V(\phi)  ;\n \\ && P = P_{\phi} = \big(\frac{\gamma -1}{\gamma}\big)f(\phi) - V(\phi)
\label{5}
\een
where the factor $\gamma$ can be mapped with the relativistic Lorentz factor \cite{Martin}:
\ben
\gamma = \frac{1}{\sqrt{1- \frac{2X}{f(\phi)}}}.
\label{6}
\een
Note that here the Lorentz factor depends on both the kinetic term ($2X$) as well as the tension scalar ($f(\phi)$). This highlights the non trivial coupling between the scalar field's dynamics and the effective geometry. In this geometry, the kinetic term governs the field’s energy density, while the tension scalar modulates the maximum attainable speed, thereby determining how relativistic effects can affect the evolution of the universe. The statement can be interpreted as follows: as the kinetic term increases, the kinetic energy of the field increases, pushing $\gamma$ towards a larger value, but the tension scalar ($f(\phi)$) sets an upper limit to the kinetic term ($2X$). In particular, for the square root in the Lorentz factor in the denominator to remain real and positive, the quantity $\frac{2X}{f(\phi)} < 1$ i.e, $\gamma >1$ must hold. Therefore, $f(\phi)$ effectively controls how fast the field can evolve, analogous to how the speed of light limits the velocity of particles in special relativity.

Now, to characterize the evolution of this scalar field, we introduce the \emph{equation of state parameter ($w_{\phi}$)} and sound speed ($c_s$), defined as:

\ben
w_{\phi} && = \frac{P_{\phi}}{\rho_{\phi}}  = \frac{ \big(\frac{\gamma -1}{\gamma}\big)f(\phi) - V(\phi)}{(\gamma - 1) f(\phi) +V(\phi)} \n \\ && =  \frac{ \big(\frac{\gamma -1}{\gamma}\big)\frac{f(\phi)}{V(\phi)} -1}{(\gamma - 1) \frac{f(\phi)}{V(\phi)} + 1} \n \\  
c_s^2 && = \frac{\frac{\partial P}{\partial X}}{\frac{\partial \rho}{\partial X}} = \frac{1}{\gamma^2}.
\label{7}
\een
The Lorentz-like factor $\gamma$ plays a central role in determining the dynamics of the DBI scalar field, as it tightly couples with $w_{\phi}$ and $c_s^2$. To facilitate the analysis, let us define the dimensionless function $ r(\phi) \equiv \frac{f(\phi)}{V(\phi)} $, since $f(\phi)$ and $V(\phi)$ have the same mass dimensions {\it [viz.} Appendix \ref{F}]. Examining the structure of the equation of state parameter (Eq.~\ref{7}), it becomes clear that $r(\phi)$ governs the overall behavior of $w_{\phi}$. Since $\gamma > 1$ by definition, even for arbitrarily large $\gamma$, the quantity $\frac{\gamma - 1}{\gamma} < 1$ always remains bounded. For the equation of state to exhibit dark energy-like behavior driving cosmic acceleration, the following inequality must be satisfied: $r(\phi)\frac{\gamma - 1}{\gamma} < 1 $. This condition ensures that the pressure remains sufficiently negative. Consequently, the functional forms of $f(\phi)$ and $V(\phi)$ must be chosen judiciously so that the above inequality holds throughout the relevant cosmological evolution.

The magnitude of $\gamma$ has direct implications for structure formation through its control of the sound speed, $c_s^2 = \frac{1}{\gamma^2}$.
A large Lorentz factor implies $c_s^2 \ll 1$, which slows the propagation of scalar perturbations and allows dark energy to cluster efficiently on sub-horizon scales, provided the background equation of state remains close to $w_{\phi} \simeq -1$. This is a distinctive feature of DBI dark energy where it can mimic a cosmological constant at the background level while permitting significant inhomogeneities at the perturbative level, which potentially leave observable signatures in the CMB and large scale structure.

K-essence models have been proposed as a compelling framework for explaining the accelerated expansion of the universe \cite{Picon1, Picon2}. The factors for this accelerated expansion include one of the possibilities is dark energy, which is effectively characterized by a non canonical scalar field, whose dynamics may be governed by the action presented in Eq. (\ref{1}). This scalar field naturally exhibits the negative pressure required for late time cosmic acceleration, thereby mimicking the behavior of dark energy. Consequently, to achieve a unified cosmological description, it is essential to include a separate matter sector alongside the scalar field. This additional sector accounts for the distribution and clustering of matter, effectively representing the role of dark matter in structure formation. Therefore, the scalar field (\ref{1}) contributes primarily to the late-time acceleration, whereas the standard pressure-less matter component accounts for structure formation in the early universe. Hence, the total action is given by:

\ben
S =S_{DBI}^{I}+ S_m
\label{8}
\een
with $S_m$ being the matter sector action, which is expressed in terms of the matter Lagrangian $\mathcal{L}_m$ as: $S_m = \int d^4x \sqrt{-g} \mathcal{L}_m$, corresponding energy momentum tensor is: $T_{\nu}^{\mu(m)}= (\rho_m+P_{m})u^{\mu}u_{\nu}- P_m\delta^{\mu}_{\nu}$. Consequently, we define an effective equation of state parameter ($\o_{eff}$) in relation to the total Lagrangian $\La =\La_m +\La_{\phi}$ later in Eq. (\ref{24a}). Henceforth, all our subsequent calculations and analyses for Model I are based on the total action defined in Eq. (\ref{8}), which encapsulates the contributions from both the k-essence scalar field and the matter component in a unified framework.\\

{\bf Model-2:} An alternative form of the DBI type scalar field Lagrangian \cite{Padmanabhan, Copeland, Sen} can be expressed as:
\ben
S_{DBI}^{II}= - \int d^4x \sqrt{-g} \big[ V(\phi)\sqrt{1- 2X}\big]
\label{9}
\een
This specific form of the action (\ref{9}) is often referred to as a tachyon-type scalar field model, inspired by developments in string theory, where tachyon condensates emerge in the context of unstable D-brane systems \cite{Sen, Padmanabhan1}. The non-canonical square root kinetic structure leads to a rich dynamical behavior that makes it a compelling candidate for modelling late time cosmic acceleration \cite{Padmanabhan, Copeland, Bagla, Scherrer, Panda, Vikman}. A detailed discussion of the mass dimensions of the scalar field $\phi$, $V(\phi)$, and the kinetic term $X$ is provided in [{\it viz.} Appendix~\ref{F}]. Following a similar procedure as before, we vary the action (\ref{9}) concerning the scalar field $\phi$ and the gravitational metric to derive the corresponding EoM and the energy density of the scalar field. In a spatially flat FLRW background, these expressions take the form:

\ben
\Ddot{\phi}+ 3\frac{H}{\gamma^2}\dot{\phi}+\frac{V'(\phi)}{V(\phi)}\frac{1}{\gamma^2}=0
\label{10}
\een
with
\ben
&&\rho_{\phi}= 2X\mathcal{L_X}-\mathcal{L}= \frac{V(\phi)}{\sqrt{1-2X}} ~;\n \\ && P_{\phi} = \mathcal{L} = - V(\phi)\sqrt{1-2X} 
\label{11}
\een
where $\mathcal{L}= -V(\phi)\sqrt{1-2X}$ and $\gamma= \frac{1}{\sqrt{1-2X}}$. Notably, the Lorentz factor ($\gamma$) is only dependent on the kinetic factor; it remains positive and real when $2X<1$.



For this Model (\ref{9}), $w_{\phi}$ and $c_s^2$ are can be expressed as:

\ben
&&w_{\phi}= \frac{P_{\phi}}{\rho_{\phi}} = -(1-2X) \n \\ && c_s^2 =\frac{\frac{\partial P}{\partial X}}{\frac{\partial \rho}{\partial X}}= \frac{1}{\gamma^2} = 1-2X
\label{14}
\een

An important observation from Eq.~\ref{14} is that, although the energy density and pressure (Eq.~\ref{11}) depend explicitly on the potential $V(\phi)$, the effective equation of state parameter $w_{\phi}$ and the sound speed $c_s^2$ depend \emph{solely} on the kinetic term $X$. This indicates that the dynamical behavior of the system is governed entirely by the kinetic structure of the scalar field rather than by its potential energy configuration.

Moreover, Eq.~\eqref{14} reveals a direct algebraic coupling between the dark energy equation of state and the sound speed via $w_{\phi} \;=\; -\, c_s^2 $. This tight relation has significant physical implications. In order for a dark energy model to cluster effectively on sub-horizon scales, two conditions are typically required: 
(i) a background equation of state close to $w_{\phi} \simeq -1$ to drive cosmic acceleration, and 
(ii) a small sound speed $c_s^2 \ll 1$ to allow dark energy perturbations to grow. However, in this purely kinetic framework these two requirements are mutually incompatible, since pushing $w_{\phi} \to -1$ forces $c_s^2 \to 1$, and conversely $c_s^2 \ll 1$ implies $w_{\phi} \simeq 0$. As a result, such models cannot simultaneously reproduce both a cosmological constant like background evolution and significant dark energy clustering. This limitation is a generic feature of the tachyonic DBI model.

Note that tachyonic DBI can act as a unified model where it can behave as dark matter and dark energy in an appropriate limit \cite{Scherrer, Padmanabhan}. However, in the context of cosmological data analysis, and considering the total matter content, which includes both dark matter and ordinary (baryonic) matter, we incorporate a standard Lagrangian for a pressureless fluid, denoted as $\mathcal{L}_m$, into the total action as follows:
\ben
S = S_{DBI}^{II} + S_m
\label{14a}
\een
where $S_m = \int d^4x \sqrt{-g} \mathcal{L}_m$.

From this point onward, we shall employ the total Lagrangian introduced above (\ref{14a}) as the basis for our cosmological data analysis of Model II. At this point, it should be mentioned that the Planck collaborations' results \cite{Planck1, Planck2, Planck3, Planck4, Planck5} observationally investigated measures of different cosmological parameters and their constraints via the DBI type action in k-essence geometry.

It is important to note that Model I in Eq. (\ref{1}) and Model II in Eq. (\ref{9}) represent two distinct classes of DBI type k-essence Lagrangian that yield the correct equation of motions mentioned in Eqs. (\ref{2}) and (\ref{10}), respectively. The primary distinction between the models lies in the functional structure and dynamical role of the scalar field kinetic term and potential. While for Model I both $f(\phi)$ and $V(\phi)$ explicitly influence the dynamics, however, in Model II, the Lagrangian depends only on the potential coupled directly to the DBI  kinetic term. It should be noted that although there is a broad freedom in choosing the functional form of $f(\phi)$ and $V(\phi)$, we adopt a specific choice that satisfies the necessary stability condition, which enables us to carry out consistent observational data analysis.\\

The term \emph{k-essence geometry} reflects the fundamentally dynamical nature of k-essence - its stress-energy tensor sources Einstein's equations within the standard framework of General Relativity, and it can backreact on the spacetime metric. The nonlinear kinetic structure of the Lagrangian thereby induces an \emph{emergent effective metric} governing the causal propagation of scalar perturbations. This emergent geometry is characterized by a generally non-luminal (dynamical) sound speed,$c_s^2 = \frac{\mathcal{L}_X}{\mathcal{L}_X + 2X\mathcal{L}_{XX}}\neq 1,$ see e.g.~\cite{Vikman, Babichev1, gm2, gm3, gm4, gm5, gm6, gm7, gm8}. Although we restrict to a spatially flat FLRW background, this emergent geometry is still encoded in the background equations of motion obtained by varying the DBI action~\cite{Vikman, Babichev1, gm3, gm4}. Hence, its imprint is already present at the level of background dynamics, and we use the term in this sense.

We will now proceed to data analysis with \PantheonShoes{} \cite{Brout, Scolnic}, \textit{Hubble} \cite{Zhang, Simon, Moresco1, Moresco2, Ratsimbazafy, Stern, Borghi, Moresco3} and BAO data \cite{Desi}. The \PantheonShoes{} dataset combines the \PantheonShoes{} compilation of Type Ia supernovae with the SHOES team's precise local measurements of the Hubble constant. It provides one of the most comprehensive and high precision datasets for constraining cosmological models and studying the universe's expansion history. The Hubble dataset consists of direct measurements of the Hubble parameter $H(z)$ at various redshifts, obtained from cosmic chronometers and BAO (Baryon Acoustic Oscillation) observations. It provides crucial information about the expansion rate of the universe over time, helping to constrain dark energy and cosmological models. The BAO  dataset captures the imprint of early universe sound waves in the large scale structure of galaxies. It serves as a standard ruler to measure cosmic distances and expansion history, providing strong constraints on dark energy and cosmological parameters.

\section{Cosmological Evolution and Observational Constraints in DBI-type k-essence Models}

The \PantheonShoes{} dataset comprises 1701 light curves corresponding to 1550 distinct Type Ia supernovae (SNe Ia), spanning a redshift range of $z=0.00122$ to $z=2.2613$ \cite{Brout, Scolnic}. However, to omit the peculiar velocity basing, we use only those data ranging from $z=0.01$ to $z=2.2613$. This includes 1590 data points (SNe Ia). The model parameters are constrained by comparing the observed and theoretical values of the distance modulus, which is defined as,

\ben
\mu(z,\theta)= m - M = 5 log_{10}(d_l(z))+25
\label{15}
\een
where $m$ is the apparent magnitude, which is how bright an object appears from Earth, and $M$ is the absolute magnitude, that is, how bright the object would appear if it were at a standard distance of 10 parsecs, and  $d_l(z)$ is the luminosity distance in Mpc, which is defined as \cite{Weinberg1}:
\ben
d_l(z)=(1+z)c\int_0^z \frac{dz}{H(z)}.
\label{16}
\een
By taking the derivative with respect to $z$, we can obtain a differential equation for $d_l(z)$ as:
\ben
\frac{dd_l(z)}{dz} && = \frac{dl}{(1+z)} + \frac{c (1+z)}{H}
\label{17}
\een
where $c$ is the speed of light measured in unit of \textit{km/s}.\\

BAO serves as a powerful probe in contemporary cosmology, enabling precise measurements of the universe’s expansion history \cite{Eisenstein}. Interpreting BAO signatures in large-scale galaxy surveys requires the use of various cosmological distance measures, namely the transverse comoving distance ($D_M$), the volume-averaged distance ($D_V$), and the Hubble distance ($D_H$). These quantities translate angular and redshift separations observed in the sky into physical scales, facilitating the extraction of key cosmological parameters such as the Hubble constant ($H_0$, present value) and the dark energy equation of state $w(z)$.

In spectroscopic galaxy surveys, the BAO feature manifests both along and across the line of sight. In the radial (line of sight) direction, the extent of the BAO peak in redshift space, denoted by a characteristic interval $\Delta z$, provides a direct measure of the Hubble parameter via
$H(z)=\frac{c\Delta z}{r_d}$ \cite{Eisenstein, Blake2, Percival},
where $r_d$ is the comoving sound horizon at the drag epoch. This relation effectively yields the Hubble distance at a given redshift, encapsulating the expansion rate of the universe at that epoch. Therefore, it measures the Hubble distance at redshift $z$:

\ben
D_H(z)=\frac{c}{H(z)}.
\label{18}
\een

In the transverse direction, the BAO scale subtends an angle $\Delta \theta$ on the sky, related to the comoving sound horizon $r_d$ through the relation $r_d=D_M(z) \Delta \theta$. By measuring this angular separation at a given redshift, one can infer the comoving angular diameter distance $D_M(z)$, which encapsulates the integrated expansion history of the universe. This distance is expressed as \cite{Eisenstein, Blake2, Percival}: 
\ben
D_M(z) = c(1 + z) \int_0^z \frac{dz'}{H(z')},
\label{19} 
\een 
where $H(z)$ is the Hubble parameter and $c$ is the speed of light. Here, the sound horizon $r_d$ serves as a standard ruler for BAO measurements. It can be expressed as \cite{Hu, Eisenstein1}:

\ben
r_d=\int_{z_{drag}}^{\infty}\frac{c_s}{H(z)}dz.
\label{20}
\een
Here, it $z_{drag}$ denotes the baryon drag epoch, the redshift at which baryons decoupled from the photon-baryon plasma. The sound horizon at this epoch, 
$r_d$ is determined by the sound speed $c_s$ of the photon-baryon fluid prior to decoupling. However, in our analysis, we treat $r_d$ as a free parameter, allowing it to be constrained directly by observational data.

When accounting for the cosmological dependence of $r_d$, the observables directly constrained by BAO measurements are typically the dimensionless ratios $\frac{D_M(z)}{r_d}$ and $\frac{D_H(z)}{r_d}$. Historically, BAO results have often been encapsulated in a single, spherically averaged distance scale \cite{Eisenstein, Blake2, Percival}, defined as: 
\ben 
D_V(z) = \left[z D_M^2(z) D_H(z)\right]^{1/3}. 
\label{21} 
\een 
This isotropic distance measure is especially useful at low redshifts, where the distinction between radial and transverse modes is less pronounced. In such Models, $D_V(z)$ provides a convenient and effective way to capture the combined geometric information from both directions in a single observable.

\subsection{Evolution equation for Model I}
To proceed with data analysis employing the total action of the form in Eq. (\ref{8}), we adopt the specific form of the tension scalar or kinetic coupling factor $f(\phi) = \lambda \phi^{4}$ and potential $V(\phi) = \frac{m^{2}\phi^2}{2}$, where $\lambda$ and $m$ are free, positive, and real parameters  \cite{Tong, Tong1, Roy}. In the framework of k-essence cosmology, the choice of the particular form of $f(\phi)$ ensures the correct DBI  dynamics, capturing strong coupling effects and enabling slow roll inflation via the ``D-cceleration" mechanism \cite{Tong}. This kinetic coupling, along with the chosen potential, $V(\phi)$ plays a crucial role in shaping the dynamics of the scalar field and its contribution to the background expansion, which may lead to rich cosmological behavior such as transitions from early to late time accelerated phase \cite{Chiba, Picon3}. Note that the appropriate dimensions of $\l$ , $m$ and $\phi$ are specified in [{\it viz.} Appendix \ref{F}]. 

In our model, $ f(\phi)$ provides a field-dependent modification to the kinetic energy of the scalar field, thereby functioning as a dynamic ``effective mass" term. This coupling allows the field's evolution to accelerate or decelerate depending on its value. By choosing the fourth power of $\phi$ form for $f(\phi)$, the system's dynamics become highly sensitive to small fluctuations in $\phi$, which offers an intrinsic mechanism to regulate the speed and characteristics of cosmic evolution.

On the other hand, the quadratic potential $V(\phi)=\frac{m^2\phi^2}{2}$ represents one of the simplest and most well studied potentials in scalar field cosmology. It introduces a mass scale for the field and supports coherent oscillatory behavior, which can mimic pressure less matter in certain regimes or drive acceleration in others, depending on the kinetic structure of the theory \cite{Unnikrishnan}. When combined with the kinetic coupling, this configuration produces a model that is both analytically tractable and dynamically rich. It supports a wide range of cosmological phenomena, making it an effective framework for studying the interaction between kinetic and potential energy in the universe's evolution.

Corresponding to the total action provided in Eq. (\ref{8}), we can write the first and second Friedmann equations as:
\ben
3H^2 &&= \rho_m + \rho_{\phi} \n \\ && =\rho_m +  (\gamma - 1) f(\phi) +V(\phi) 
\label{22}
\een
and,
\ben
\frac{\Ddot{a}}{a} &&= -\frac{1}{6}(\rho_m + 3P_m + \rho_{\phi}+ 3P_{\phi})\n \\ && = -\frac{1}{6}\big(\rho_m +\rho_{\phi}(1+3w_{\phi})\big)
\label{23}
\een
where the matter sector energy density ($\rho_m$), and the dark energy sector energy density ($\rho_{\phi}~(\equiv \rho_{d})$) satisfy the following differential equation: 
\ben
&&\frac{d\rho_m}{dt} = - 3 H\rho_m \n \\ && \frac{d\rho_{\phi}}{dt} = - 3 H \rho_{\phi}(1+ w_{\phi})
\label{24}
\een
In this framework, we model matter as pressureless ($P_{m}=0$), i.e., $w_m = 0$, while $w_{\phi}~(\equiv w_{d}\equiv w_{de})$ is given by Eq.~(\ref{7}). We further introduce an {\it effective} equation-of-state parameter, $\omega_{\text{eff}}~(\equiv\frac{P_{eff}}{\rho_{eff}})$, as defined with respect to in Eq.~(\ref{8}), by setting $\rho_{\text{eff}} = \rho_m + \rho_{\phi}$ and $P_{\text{eff}} = P_{\phi}$. Using Eqs.~(\ref{22}) and (\ref{23}), $\omega_{\text{eff}}$ can then be written in terms of the redshift $z$ and the Hubble parameter.:
\ben
\o_{eff} = -1 -\frac{2}{3}\frac{\frac{dH}{dt}}{H^2} = -1 +\frac{2 (1+z)}{3}\frac{\frac{dH}{dz}}{H}.
\label{24a}
\een

The above Eq. \ref{24} for $\rho_m$ and $\rho_{\phi}$ can be rewritten in terms of the redshift $z$ as

\ben
&&\frac{d\rho_m}{dz} = \frac{d\rho_m}{dz}\frac{dz}{dt} =   \frac{3 ~ \rho_m}{(1+z)};\n \\ &&
\frac{d\rho_{\phi}}{dz} = \frac{d\rho_{\phi}}{dz}\frac{dz}{dt}= \frac{3\rho_{\phi}(1+ w_{\phi})}{1+z}
\label{24b}
\een

We now define the {\it dimensionless dark energy density parameter ($\O_{\phi}\equiv\O_{d}$) and dimensionless total matter (i.e., dark matter and ordinary matter) density parameter ($\O_{m}$)} using Eq. (\ref{22}) as \cite{Weinberg1}:
\ben
\Omega_{\phi}\equiv \Omega_{d}=\frac{\rho_{\phi}}{3H^2}; ~ \Omega_{m}= \frac{\rho_{m}}{3 H^2}
\label{25}
\een
the evolution equation for $\O_m$ corresponding to redshift distance ($z$) can be written as:
\ben
\frac{d\Omega_{m}}{dz} = \frac{\Omega_{m}}{1+z}\Big[3 -2\frac{(1+z)}{H}\frac{dH}{dz}\Big]
\label{26}
\een
where we can express $\frac{dH}{dz}$ using Eqs. (\ref{22}), (\ref{23}), (\ref{24}) as:
\ben
\frac{dH}{dz} = \frac{3 H}{2(1+z)}\big(\Omega_{d}w_{\phi} +1\big).
\label{27}
\een

As observationally redshift distance ($z$) is the crucial parameter rather than time ($t$), we must express all equations governing the evolution of Hubble parameter ($H$), density parameters ($\O_d$ or $\O_m$), scalar field ($\phi$), and field velocity ($v = \frac{d\phi}{dt}$) in terms of $z$. The fundamental relation between cosmic time and redshift follows from $a = (1+z)^{-1}$. Differentiating with respect to $t$ and using $H \equiv \dot{a}/a$ gives
\ben
&&\frac{dz}{dt} = -(1+z)\,H(z), \n \\ && \frac{dt}{dz} = -\frac{1}{(1+z)\,H(z)}.
\label{27a}
\een
This relation is used throughout to convert time derivatives into redshift derivatives. For any quantity, namely, $Q(t)$: $dQ/dz = \dot{Q}\cdot dt/dz = -\dot{Q}/[(1+z)H(z)]$. Therefore, we can express the scalar field EoM (\ref{2}) in terms of redshift distance $z$ as:
\ben
\frac{d\phi}{dz} &&=\frac{d\phi}{dt}\frac{dt}{dz} = - \frac{v}{H(1+z)} \n \\
\frac{dv}{dz} &&= - \frac{3H}{\gamma^2}\frac{d\phi}{dz} - \frac{3f'(\phi)}{2f(\phi)}\frac{v^2}{H(1+z)} +\frac{f'(\phi)}{H(1+z)}(1-\frac{1}{\gamma^3})\n \\ && +\frac{1}{\gamma^3}\frac{V'(\phi)}{H~(1+z)},
\label{28}
\een

Now, to facilitate our analysis, we make use of the square root kinetic structure of Model I~(\ref{8}). For a more convenient dynamical formulation, we define the dimensionless variable
$x \equiv \frac{v}{\sqrt{\lambda}\,\phi^{2}}, \quad v \equiv \dot{\phi}=\frac{d\phi}{dt}$, which recasts the scalar field evolution into a compact first-order system. The full background equations can then be rewritten in terms of the redshift variable $z$. This reformulation allows the coupled dynamics of the DBI field and the cosmological background to be expressed in a form well suited for numerical integration and parameter estimation. Using Eq. (\ref{28}), we can write the differential equation satisfied by the scalar field $\phi$ and our dimensionless variable $x$ by explicitly putting the form $f(\phi)$ and $V(\phi)$ as:

\ben
&&\frac{d\phi}{dz} = - \frac{x \sqrt{\l} \phi^2}{H(1+z)} ; \n \\ 
&& \frac{dx}{dz} =  \frac{1}{H(1+z)}\Big(\frac{3 H x}{\gamma^2} + 4 \sqrt{\l} \phi \big(1 - \frac{1}{\gamma^3} - x^2\big) \n \\ &&+ \frac{m^2}{\sqrt{\l} \phi \gamma^3}\Big) 
\label{28a}
\een
where we define $\gamma = \frac{1}{\sqrt{1 - x^2}}$ (\ref{6}).

\subsection{Evolution equations for model II}

For the second DBI type k-essence model (\ref{14a}), we take the form of the potential coupled to the non-canonical kinetic term as $V(\phi) = \frac{m^{2}\phi^2}{2}$, where we again choose $m$ to be a free, positive, and real parameter. The quadratic potential, a simple non linear potential, is a suitable choice for modeling the dark sector of the universe. The quadratic type potential is also a slow roll compatible potential in standard inflation \cite{Mukhanov1} and allows analytic progress in the tachyonic type DBI  Model as well \cite{Sen}. Note that, in this analysis too, dimensional consistency has been ensured while the explicit determination of the actual dimension is provided in [{\it viz.} Appendix \ref{F}]. 

For the total action in Eq.~\ref{14a}, the total energy density naturally splits into two contributions: the pressure-less matter component arising from $\mathcal{L}_m$, denoted by $\rho_m$, and the dark energy component originating from the DBI sector in Eq.~\ref{9}, denoted by $\rho_{\phi}$. With this decomposition, the background evolution of the universe continues to be governed by the Friedmann equations, given in Eqs.~(\ref{22}), (\ref{23}), however, with the modified definition of $\rho_{\phi}$ and $P_{\phi}$ as defined in Eq. (\ref{11}). However, the continuity equation retains the form of Eq.~(\ref{24}). Using the same definitions for the matter and dark energy density parameters, we straightforwardly recover the Eqs.~(\ref{25})–(\ref{27}) as dimensionless energy densities. Consequently, the evolution equation, together with the same reasoning applied earlier (prior to Eq.~(\ref{28})) enables us to rewrite the EoM (\ref{10}) in terms of the redshift $z$ as:

\ben
&&\frac{d\phi}{dz}=\frac{d\phi}{dt}\frac{dt}{dz} = - \frac{v}{H(1+z)} \n \\
&&\frac{dv}{dz}= -3\frac{H}{\gamma^2}\frac{d\phi}{dz} + \frac{1}{H(1+z)}\frac{1}{\gamma^2}\frac{V'(\phi)}{V(\phi)}.
\label{33}
\een 

In our method for models I \& II, obtaining the evolution equations used for parameter inference via Bayesian analysis within an ML framework, it is equally valid to infer either dark energy ($\Omega_d$) or dark matter ($\Omega_m$) density parameter and the choice can be made without loss of generality. The two parameters are intrinsically linked by the constraint $\Omega_{d} + \Omega_{m} =1$, which holds under the assumption of a spatially flat universe and in the absence of significant contributions from radiation or other exotic components at low redshift. It is worth noting that in principle, a complete energy budget of the universe should include a radiation component characterized by the density parameter $\O_R$. However, in the present epoch, the contribution from radiation, comprising photons and relativistic neutrinos, is exceedingly small compared to that of dark matter and dark energy. Owing to the substantial expansion of the universe since the radiation dominated era, the current value of $\Omega_R$ is approximately $10^{-5}$ \cite{Planck3}, rendering its influence negligible in the context of late time cosmological dynamics. Consequently, we omit the radiation component in our analysis for late time. 

Therefore, determining the evolution of either $\Omega_d(z)$ or $\Omega_m(z)$ at late times is sufficient to fully specify the other. In practice, it is often more convenient to work with the total matter density $\Omega_m(z)$, since it follows directly from the continuity equation (\ref{24}) or (\ref{24b}). By contrast, the evolution of dark energy in models with dynamical or non-standard equations of state depends explicitly on the underlying scalar-field dynamics or on a chosen phenomenological parametrization, such as the Chevallier--Polarski--Linder (CPL) \cite{Chevallier}. In this work, we employ the CPL parametrization (see \emph{Appendix \ref{D}}) as a benchmark against which our k-essence models are compared.\\

Another point to mention is that, in the context of scalar field cosmology, particularly for models like DBI-type k-essence ((\ref{1}) or (\ref{8}) and (\ref{9}) or (\ref{14a})) the use of differential equations to determine the evolution of the Hubble parameter $H(z)$, the scalar field $\phi(z)$, and its velocity $v(z)=\frac{d\phi}{dt}$, are essential for ensuring a self consistent and physically accurate description of the dynamics. These quantities are intricately coupled: the evolution of $H(z)$ depends on the energy content of the universe, which in turn is governed by the scalar field dynamics, while $\phi(z)$ and $v(z)$ evolve according to equations that depend explicitly on $H(z)$ and $\gamma(z)$, and the scalar field potential ($V(\phi)$). Solving the full set of differential equations allows the system to evolve naturally from initial conditions, preserving the interdependence dictated by the underlying theory. Moreover, in the context of data analysis and parameter inference, evolving the system via differential equations guarantees that predictions are consistently tied to the model parameters, enabling robust comparison with observational data. 

A key point concerns the consistent treatment of units across the theoretical and observational parts of our analysis. Throughout the theoretical derivation, including the Friedmann equations~(\ref{22})--(\ref{23}), the continuity equation~(\ref{24}), and the scalar-field evolution equations~(\ref{28}), (\ref{28a}), and (\ref{33}), we work in reduced Planck units with $\hbar = c = 1$ and $M_{\rm Pl} \equiv (8\pi G)^{-1/2} = 1$. As a result, factors such as $8\pi G$ do not appear explicitly in Eqs.~(\ref{22}) and (\ref{23}), since they are absorbed into the unit convention. Equivalently, in these units one has $8\pi G = 1$ identically.

Nevertheless, the physical dimension of the Hubble parameter is always retained as $[T^{-1}]$, which ensures that the scalar-field equations, continuity equations, and Friedmann equations remain dimensionally consistent. When connecting the theoretical evolution to observational quantities, we restore conventional cosmological units. In particular, the Hubble parameter is expressed in the standard observational form km\,s$^{-1}$Mpc$^{-1}$.

This is also the reason why the speed of light $c$ appears explicitly in the luminosity-distance relation Eq.(\ref{16}),
since the integral over $1/H(z)$ alone has dimensions of time, and multiplication by $c$ converts it into a physical distance. Therefore, there is no inconsistency between the theoretical equations and the observational expressions. The apparent absence of $8\pi G$ simply reflects our use of reduced Planck units in the theoretical part of the analysis.

\section{Methodology and Data Analysis with Model Fitting}

In this work, we numerically evolve a DBI-type k-essence cosmological model forward in redshift over the interval $0 \rightarrow 2.5$ using \texttt{Diffrax} \cite{Kidger}. This integration yields theoretical predictions for key observables, including the Hubble expansion rate $H(z)$, the luminosity distance $d_{L}(z)$, and BAO distance ratios. We construct a training set of $50000$ cosmological parameter realizations for both models by sampling from uniform prior distributions over $1000$ epoch. For each sampled parameter set, the background evolution is solved numerically to obtain the complete observable vector that incorporates Type Ia supernova data (\PantheonShoes{}), Hubble parameter measurements, and BAO constraints.

We employ a neural network based surrogate model, implemented in \texttt{Flax}\cite{Google} as a multilayer perceptron (MLP)  with three hidden layers of 256 neurons each and ReLU~(Rectified Linear Unit) activations \cite{Nair}, to approximate the full forward cosmological mapping of the DBI k-essence model and thus enable fast, accurate likelihood evaluations. Training is guided by validation loss monitoring to ensure stable convergence, using a cosine learning rate schedule and an early stopping rule that terminates optimization after 50 epochs without validation loss improvement. Once converged, the surrogate replaces repeated ODE integrations, yielding a substantial acceleration of the Bayesian inference pipeline.

We perform Bayesian parameter estimation via stochastic variational inference (SVI) \cite{Hoffman1} using a multivariate normal variational family implemented with the \texttt{Auto-Multivariate-Normal} guide \cite{Bingham}. The variational distribution is fitted by minimizing the negative evidence lower bound (ELBO) with the \texttt{Adam} optimizer \cite{Kingma}, employing early stopping to ensure stable convergence and mitigate over-fitting. SVI is run for up to $5\times 10^{4}$ iterations, after which the best converged variational parameters are used to approximate the posterior. Thereafter, samples from this variational posterior are propagated in redshift by numerically integrating the underlying system of ordinary differential equations, yielding the background cosmological evolution. From these reconstructed histories, we compute derived dynamical quantities, including the dark energy equation of state $w_{d}(z)$, the effective total equation of state $w_{\mathrm{eff}}(z)$, the deceleration parameter $q(z)$, and the acceleration-deceleration transition redshift $z_{t}$.

To assess model performance and to facilitate model comparison, we evaluate standard information criteria and goodness-of-fit statistics namely, $\chi^{2}$, AIC~\cite{Akaike}, BIC~\cite{Schwarz}, and DIC~\cite{Spiegel} - directly from the ensemble of posterior realizations. In addition, we calculate fully Bayesian model selection criteria, namely the Watanabe-Akaike Information Criterion (WAIC) \cite{Watanabe}, and the Pareto-smoothed Importance Sampling Leave-One-Out cross-validation (PSIS LOO) \cite{Vehtari}, for both models to obtain a comprehensive and statistically robust assessment of their performance. The resulting posterior geometry and parameter correlations are examined using \texttt{GetDist} \cite{Lewis}, providing a detailed visualization of the inferred parameter space. In addition, symbolic regression with \texttt{PySR} \cite{Boyle} is employed on the reconstructed dark energy histories to obtain compact, analytic surrogate expressions for the redshift dependence of the dark energy dynamics, thereby yielding interpretable, closed-form representations of the numerically inferred behavior. 

Overall, this constitutes a fully ML-driven, physics-informed cosmological inference framework, in which we deliberately avoid the conventional MCMC sampling paradigm. Rather than generating long MCMC chains in the traditional Bayesian analysis \cite{Gelman}, we instead train a probabilistic model that learns the mapping from observational data to the allowed region of parameter space under the constraint that the relevant cosmological evolution equations are satisfied. Once this physics-constrained mapping is established, SVI is employed to perform approximate Bayesian inference, producing a variational posterior that closely mimics the posterior distribution typically obtained via MCMC, but at substantially reduced computational cost and with a more seamless integration of ML and physical modeling.

We will also find the functional form of $w_{\phi}$  from the graph (\ref{Fig3} and \ref{Fig5}) of the EoS parameter ($w_{\phi}$) obtained from the Bayesian best fit parameter by using PySR. PySR implements symbolic regression using genetic programming (GP) to evolve explicit mathematical expressions that approximate a target function from data. It represents candidate solutions as tree structures composed of basic mathematical operations. It iteratively improves them through genetic operators like mutation, crossover, and selection, optimizing for both model accuracy and complexity.

To conduct the Bayesian analysis, we formulate a six-parameter ({\bf p}=($H_0$, $\Omega_{m0}$, $r_d$, $\lambda$, $\phi_0$, $x_0$)) differential Eqs. (\ref{17}), (\ref{28a}) along with (\ref{26}) concerning the first kind of Lagrangian corresponding to the action (\ref{8}) and additionally formulate a five-parameter ({\bf p}=($H_0$, $\Omega_{m0}$, $r_d$, $\phi_0$, $v_0$)) differential Eqs. (\ref{17}), (\ref{33}) along with (\ref{26}) for the second kind of Lagrangian corresponding to the action (\ref{14a}). The solution of these equations with the initial condition is subsequently fitted against the available data set of the type Ia supernova data (\PantheonShoes{} data) \cite{Brout, Scolnic}, \textit{Hubble} data \cite{Zhang, Simon, Moresco1, Moresco2, Ratsimbazafy, Stern, Borghi, Moresco3} and for the \textit{BAO} dataset \cite{Desi}. In our analysis, we use the latest data release from the DESI collaboration DR2 dataset \cite{Hussain2}.  We employ the $\chi^2$ statistics to constrain the model parameters, thereby assessing the discrepancy between the observed data and a theoretical model. This statistical technique is widely used in hypothesis testing and model fitting to evaluate the adequacy of a model in describing the provided data. For fitting the Bayesian model \cite{Hoffman}, $\chi^2$ is frequently incorporated into likelihood functions: $\textit{L} \propto exp(-\frac{1}{2}\chi^2)$. This establishes a connection between chi squared minimization and Maximum Likelihood Estimation (MLE), which proves beneficial for Bayesian posterior sampling.

For the models with model parameters ($\mathbf{p}$), we compute the $\chi^2$ function for the Hubble dataset given in Table \ref{TableIV} ({\it viz. Appendix}) as per \cite{Hogg, Hogg1}:
\ben
\chi_{H}^2 = \sum_{i=1}^{{\color{blue}32}} \left(\frac{H^{th}(z_i,{\mathbf{p}})-H^{obs}(z_i)}{\sigma_i}\right)^2
\label{34}
\een
where $H^{th}(z_i,{\mathbf{p}})$ represents the theoretical value obtained by solving the differential equations, and $H^{obs}(z_i)$ corresponds to the value provided in the $H(z)$ column of Table \ref{TableIV}. $\sigma_i$ denotes the error or uncertainty in the measurement of $H(z)$ as mentioned in the same table.

The $\chi^2$ for BAO data can be computed as \cite{Hogg, Hogg1}:

\ben
\chi_{i}^2 = \sum_{i=1}^{N} \left(\frac{X_{i}^{th}(z,{\mathbf{p}})-X_{i}^{obs}(z_i)}{\sigma_{i}}\right)^2,
\label{35}
\een
where $X_{i}^{th}$ is either $\frac{D_M}{rd}$, $\frac{D_H}{r_d}$, or $\frac{D_V}{r_d}$, calculated theoretically by model fitting. Therefore, the total $\chi^2_{t}$ corresponding to the BAO dataset is obtained as $\chi^2_{t} = \left(\chi^2_{(\frac{D_M}{r_d})}+\chi^2_{(\frac{D_V}{r_d})}+ \chi^2_{(\frac{D_H}{r_d})}\right)$.\\

For the PANTHEON data set corresponding to SN Ia supernovae \cite{Brout, Scolnic}, we employ a different method to obtain $\chi^2$. This is due to the presence of a covariance matrix ($C$) of dimension $1701 \times 1701$, which corresponds to the measurement error of the distance modulus ($\mu(z)$) for all 1701 light curves. The expression for $\chi^2$ is as follows:

\ben
\chi_{SN}^2 &&= \Big(\mu_{th}(z_i,{\mathbf{p}})-\mu_{obs}(z_i)\Big)^TC^{-1}\nonumber\\  &&\times\Big(\mu_{th}(z_i,{\mathbf{p}})-\mu_{obs}(z_i)\Big) 
\label{36}
\een
Here, $\mu_{th}(z_i, \mathbf{p})$ is the theoretical prediction obtained from Eq. (\ref{15}) by solving the differential equations (\ref{26})–(\ref{28}) corresponding to the action (\ref{8}), or alternatively by solving Eq. (\ref{33}) along with Eqs. (\ref{26}) and (\ref{27}) using the modified definitions of $\rho_{\phi}$ and $P_{\phi}$ given in (\ref{11}) for the action (\ref{14a}), subject to the chosen initial conditions and model parameters $\mathbf{p}$. The quantity $\mu_{obs}$ is the observed distance modulus from the PANTHEON compilation \cite{Brout, Scolnic}, and $C^{-1}$ is the inverse covariance matrix.

\subsection{Model I}

Using NumPyro to perform SVI optimization with ML based surrogate evaluations, we obtain the inferred best fit parameters together with their associated uncertainties, which are reported in Table \ref{TableI}.

Because the DBI equations of motion are highly nonlinear, the model responds very sensitively to variations in the input parameters. In practice, naive or unstructured sampling of the parameter space often produces numerical instabilities, including divergent solutions, or yields cosmological histories that fail to satisfy the flatness requirement ($\Omega_{\mathrm{tot}} \neq 1$). This strong dependence on parameter choice makes it essential to carefully delineate and justify the parameter ranges explored in our analysis, so as to avoid unphysical trajectories and ensure numerically stable evolution. 
\begin{table*}[ht]
\centering
\begin{tabular}{|c|c|c|}
\hline
\textbf{Parameter} & \textbf{Prior} & \textbf{SVI best-fit}  \\
\hline
$\Omega_{m0}$ & $\mathcal{U}(0.20, 0.40)$ & $0.307 \pm 0.011$  \\
\hline
$H_0$[Km/s/Mpc]  & $\mathcal{U}(60,80)$ & $73.29 \pm 0.16$ \\
\hline
$r_d$[Mpc] & $\mathcal{U}(120,170)$   & $138.08 \pm 0.71$ \\
\hline
$\lambda$ & $\mathcal{U}(0.001, 5.0)$  & $2.48 \pm 0.98$ \\
\hline
$\phi_0$[GeV] & $\mathcal{U}(0.001, 5.0)$  & $ 3.70^{+0.74}_{-0.42} $  \\
\hline
$x_0$ & $\mathcal{U}(-0.99, 0.0)$ & $-0.609 \pm 0.05$ \\
\hline
\end{tabular}
\caption{Parameter estimates and uncertainties using Bayesian inference with NumPyro via SVI optimization for Model I.}
\label{TableI}
\end{table*}

\begin{figure}[ht] 
    \centering
    \includegraphics[width=\columnwidth]{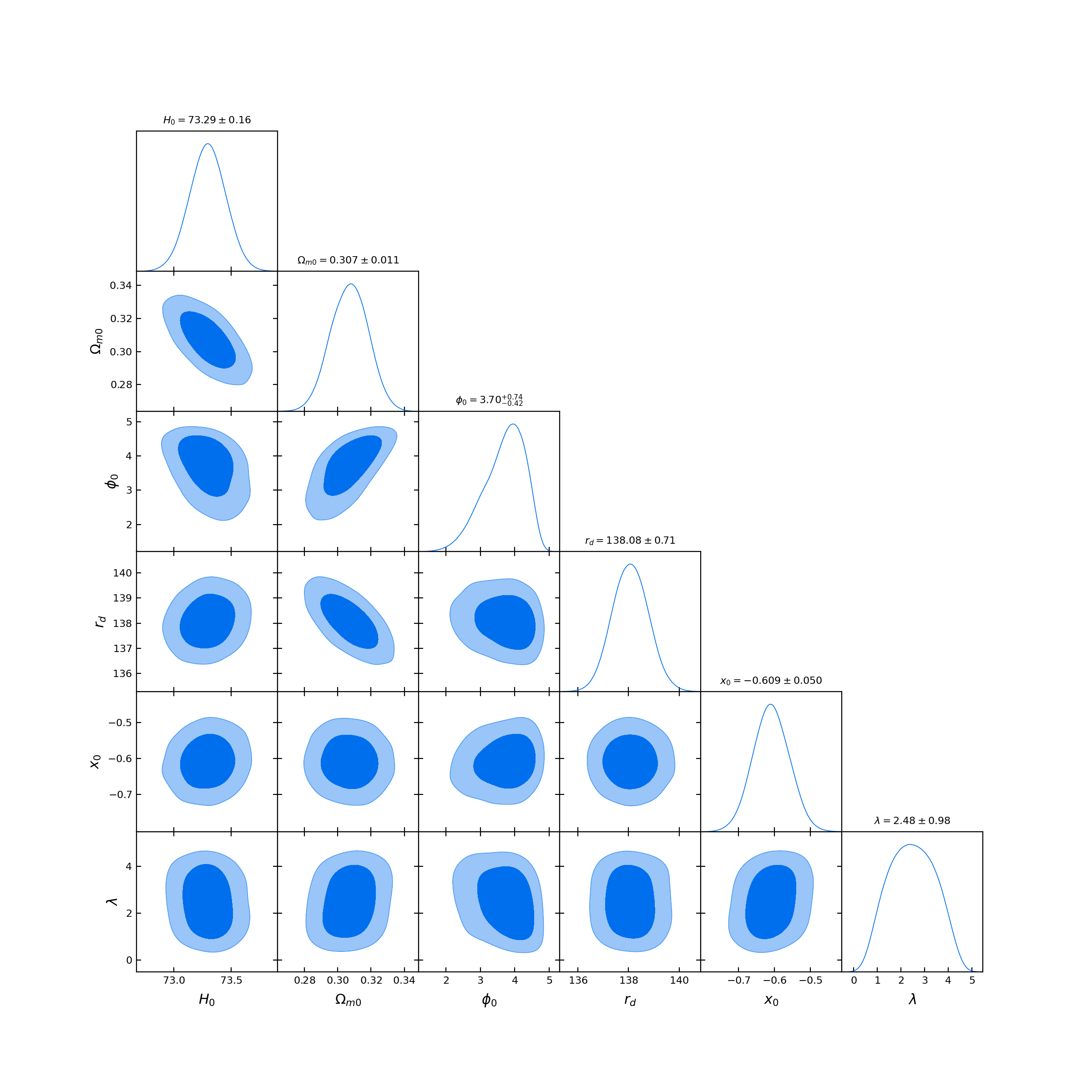}
    \caption{Bayesian inference using SVI for the combined data set of Hubble+ \PantheonShoes{} BAO for Model I}
    \label{Fig1} 
\end{figure}

Figure~\ref{Fig1} displays the joint posterior distributions associated with Model~I, derived using SVI sampling, where the likelihood evaluations are accelerated via the ML surrogate emulator. The one-dimensional marginalized posterior distributions are narrowly concentrated and exhibit a single pronounced maximum, signalling satisfactory convergence of the inference procedure and indicating that the combined dataset imposes tight constraints on the model parameters. The two-dimensional credible regions, however, expose several non-trivial degeneracies among different parameters, reflecting residual correlations in the inferred parameter space. It is important to emphasize that in this analysis we explicitly vary only six parameters ({\bf p}=($H_0$, $\Omega_{m0}$, $\phi_0$, $x_0$, $r_d$, $\l$)), excluding $m$ from the set of free parameters. The justification for treating $m$ as a fixed rather than inferred quantity is provided in detail in {\it Appendix \ref{B}}.

\subsection{Model II}
For the second model using the action given in (\ref{14a}), we additionally perform a Bayesian analysis based on SVI optimization, implemented in NumPyro and utilizing an ML-based emulator approach. This procedure enables us to identify the optimal best-fit values of the model parameters along with their associated uncertainties. The resulting parameter estimates and error ranges are summarized in Table \ref{TableII}.

\begin{table*}[ht]
\centering
\begin{tabular}{|c|c|c|c|}
\toprule
\textbf{Parameter} & \textbf{Prior} & \textbf{SVI best-fit} \\
\midrule
    $\Omega_{m0}$ & $\mathcal{U}(0.20,0.40)$  & $0.2960 \pm 0.0049$ \\
    \hline
    $H_0$[Km/s/Mpc] & $\mathcal{U}(60,80)$ & $73.35 \pm 0.15$ \\
    \hline
    $r_d$[Mpc] & $\mathcal{U}(120, 170)$ & $138.81 \pm 0.57$ \\
    \hline
    $\phi_0$[GeV] & $\mathcal{U}(0.001, 20)$ & $7.9^{+2.9}_{-4.6}$ \\
    \hline
    $v_0$[GeV/s] & $\mathcal{U}(-0.99, 0.99)$ & $0.003 \pm 0.019$ \\
\bottomrule
\end{tabular}
\caption{Parameter estimates and uncertainties using Bayesian inference with NumPyro via SVI optimization for Model II.}
\label{TableII}
\end{table*}

In this analysis, we again restrict ourselves to five parameters, {\bf p} = ($H_0$, $\Omega_{m0}$, $r_d$, $\phi_0$, $v_0$), and omit the mass parameter $m$ from the set of independent variables. Instead, we impose an additional constraint that fixes the value of $m$. Thus, $m$ is determined as a function of the four parameters (${H_0, \Omega_{m0}, \phi_0, v_0}$) and is fully specified by Eq.~(\ref{B4}), as detailed in  {\it viz.} Appendix~\ref{B}.

\begin{figure}[ht] 
    \centering
    \includegraphics[width=\columnwidth]{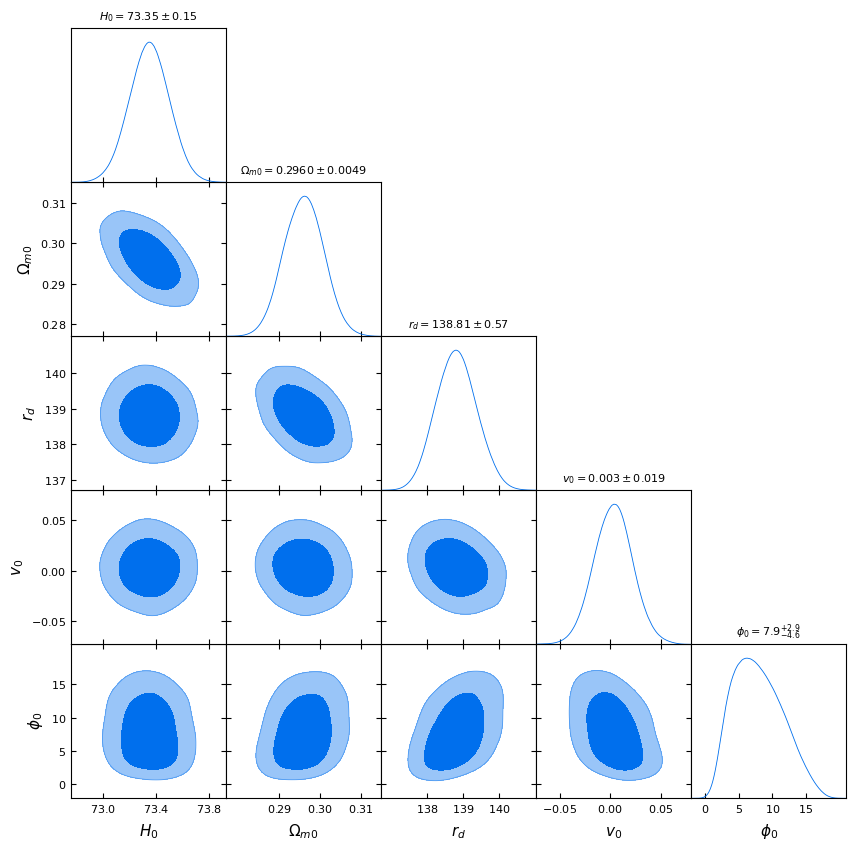}
    \caption{Bayesian analysis using SVI for the combined data set of Hubble+ \PantheonShoes{} BAO for Model II }
    \label{Fig2} 
\end{figure}

In Fig.~(\ref{Fig2}), we present the posterior probability distributions of the model parameters in both one and two dimensional representations. The one dimensional distribution exhibit approximately Gaussian, unimodal behavior, indicating a single dominant region of high probability for each parameter. In contrast, the two dimensional joint distributions reveal correlations ranging from strong to moderate among the parameters, as evidenced by their elongated and tilted credible regions in the parameter space.

Note that, in both realizations of the model, we do not treat $m$ as an independent free parameter. Instead, its value is determined implicitly through a constraint equation. The introduction of this constraint is physically motivated and is specifically designed to enforce the flatness condition $\Omega_m + \Omega_d = 1$ at all times. Because the scalar field in our framework is constructed to describe only the dark energy component, we restrict $m$ via the explicit relations provided in \ref{B1} and \ref{B3}. By imposing these relations, we ensure that the late-time cosmological dynamics remain compatible with a spatially flat universe, so that the combined contribution of the matter and dark energy density parameters is identically equal to unity throughout the entire dynamical regime of interest.

\textbf{Explanation of higher $H_0$ than reported by Pantheon$+$SH0ES compilation:} Table~\ref{TableI} and Table~\ref{TableII} give $H_0=73.29\pm0.16$\,km\,s$^{-1}$Mpc$^{-1}$ (Model~I) and $H_0=73.35\pm0.15$\,km\,s$^{-1}$Mpc$^{-1}$ (Model~II), exceeding the SHOES central value $H_0 = 73.04 \pm 1.04$\,km\,s$^{-1}$Mpc$^{-1}$ \cite{Riess1} by $\approx 0.25$--$0.31$\,km\,s$^{-1}$Mpc$^{-1}$, i.e.\ within $0.3\sigma$ and therefore statistically fully consistent. The small but systematic excess above the SH0ES central value is not a numerical artefact; it has a direct and transparent physical origin within the tachyonic DBI framework.

Both models yield a dark-energy equation of state satisfying $w_\phi \ge -1$, which \emph{increases toward less negative values with increasing redshift} (Figs.~\ref{Fig3} and~\ref{Fig5}). Quantitatively, Model~I yields $w_{d,0} \simeq -0.997\pm0.007$ at $z=0$, while the tachyonic identity $w_\phi = -c_s^2$ of Model~II enforces $w_{d,0} > -1$ whenever the kinetic term $X > 0$. The physical consequence is the following - at intermediate redshifts $z \sim 0.5$--$2$, where the supernova and BAO data are most constraining, a dark-energy component with $w_\phi > -1$ contributes \emph{less} energy density to the total expansion budget than a cosmological constant would at identical $\Omega_{m0}$. This reduction in the dark-energy contribution compresses the integrated comoving distances relative to $\Lambda$CDM, and to recover the same observed supernova distance moduli and BAO ratios the Friedmann equation must compensate through a \emph{larger} present-day expansion rate. In other words, the data trade a slightly less dominant dark energy in the recent past for a slightly faster expansion today. Therefore, the deficit in dark-energy density at intermediate redshifts is absorbed by an upward shift in $H_0$. The resulting $\Delta H_0 \approx +0.3$\,km\,s$^{-1}$Mpc$^{-1}$ excess is therefore a direct and physically motivated kinematic imprint of the $w_\phi > -1$ tachyonic DBI dynamics on the inferred Hubble constant.\\

\subsection{Evolution of the equation of state parameter for both the model:}
\begin{figure}[ht] 
    \centering
    \includegraphics[width=\columnwidth]{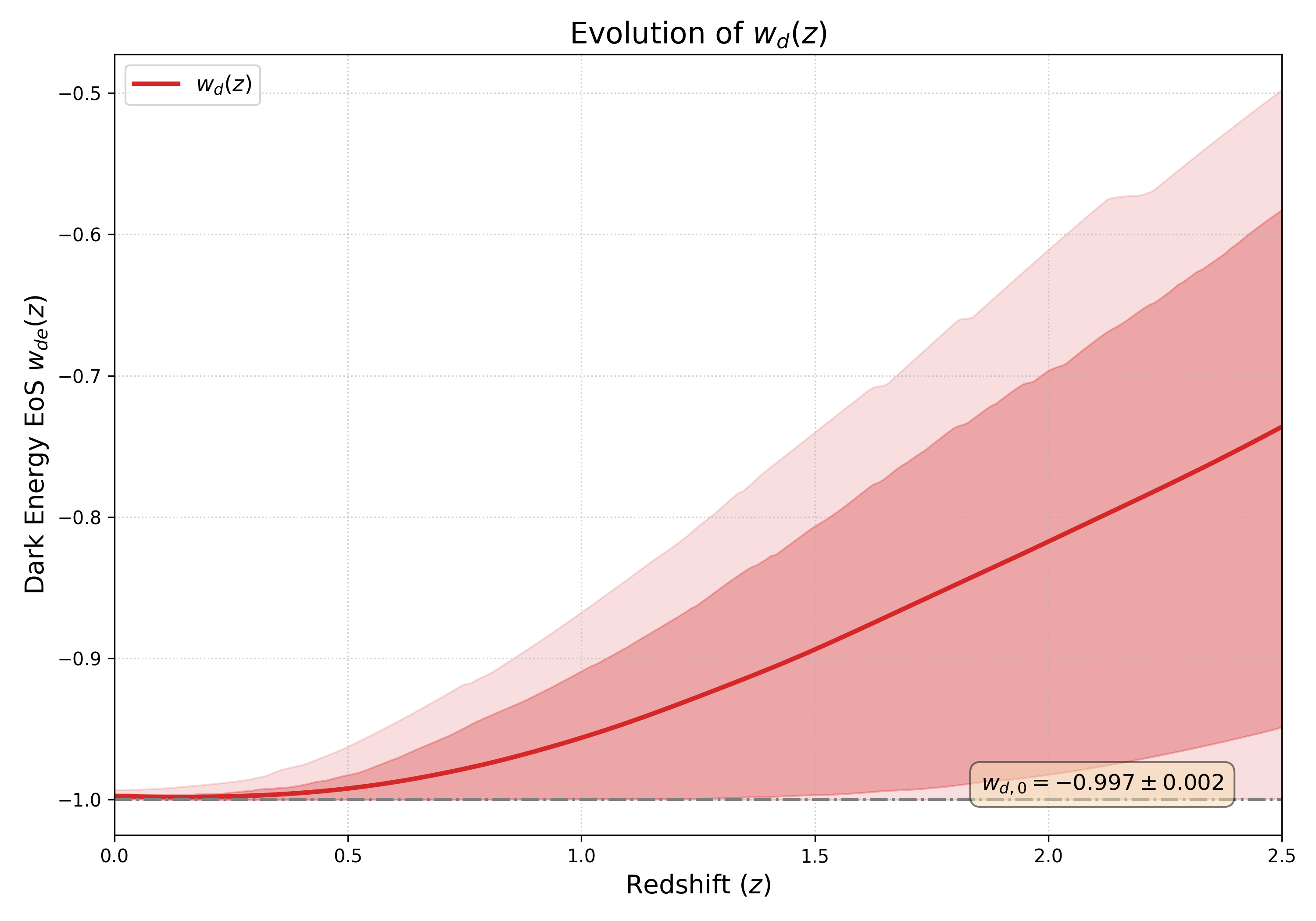}
    \caption{Plot of dark energy EoS parameter ($w_{d}$) (Eq. \ref{7}) with redshift distance ($z$) with 1$\sigma$ and 2$\sigma$ confidence interval.  The $w=-1$ line corresponding to the $\Lambda$CDM model is shown as a grey `dashed-dot' line.}
    \label{Fig3} 
\end{figure}

\begin{figure}[ht] 
    \centering
    \includegraphics[width=\columnwidth]{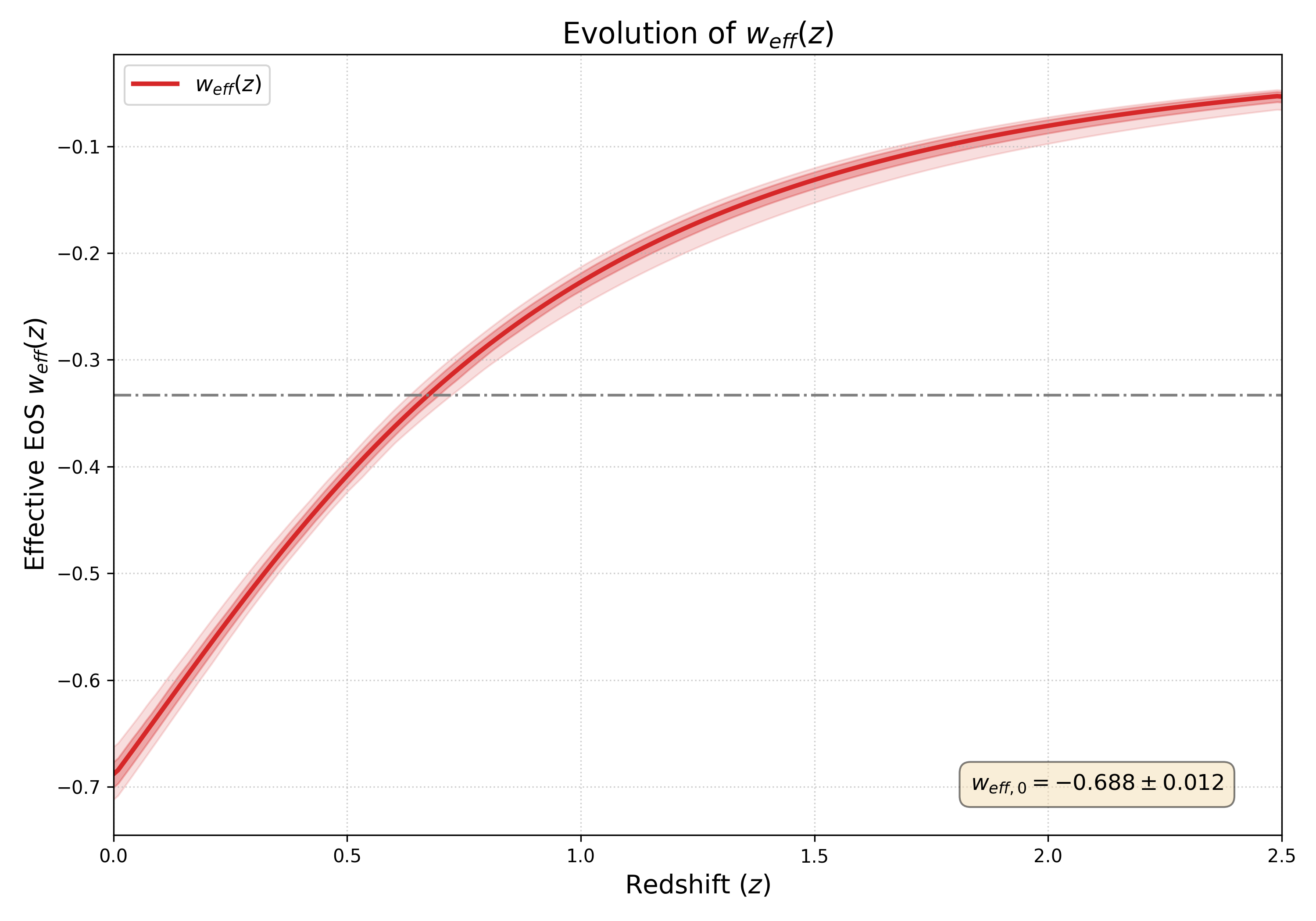}
    \caption{Plot of effective EoS parameter ($w_{eff}$) with redshift distance ($z$) for Model I with 1$\sigma$ and 2$\sigma$ confidence interval.}
    \label{Fig4} 
\end{figure}

\begin{figure}[ht] 
    \centering
    \includegraphics[width=\columnwidth]{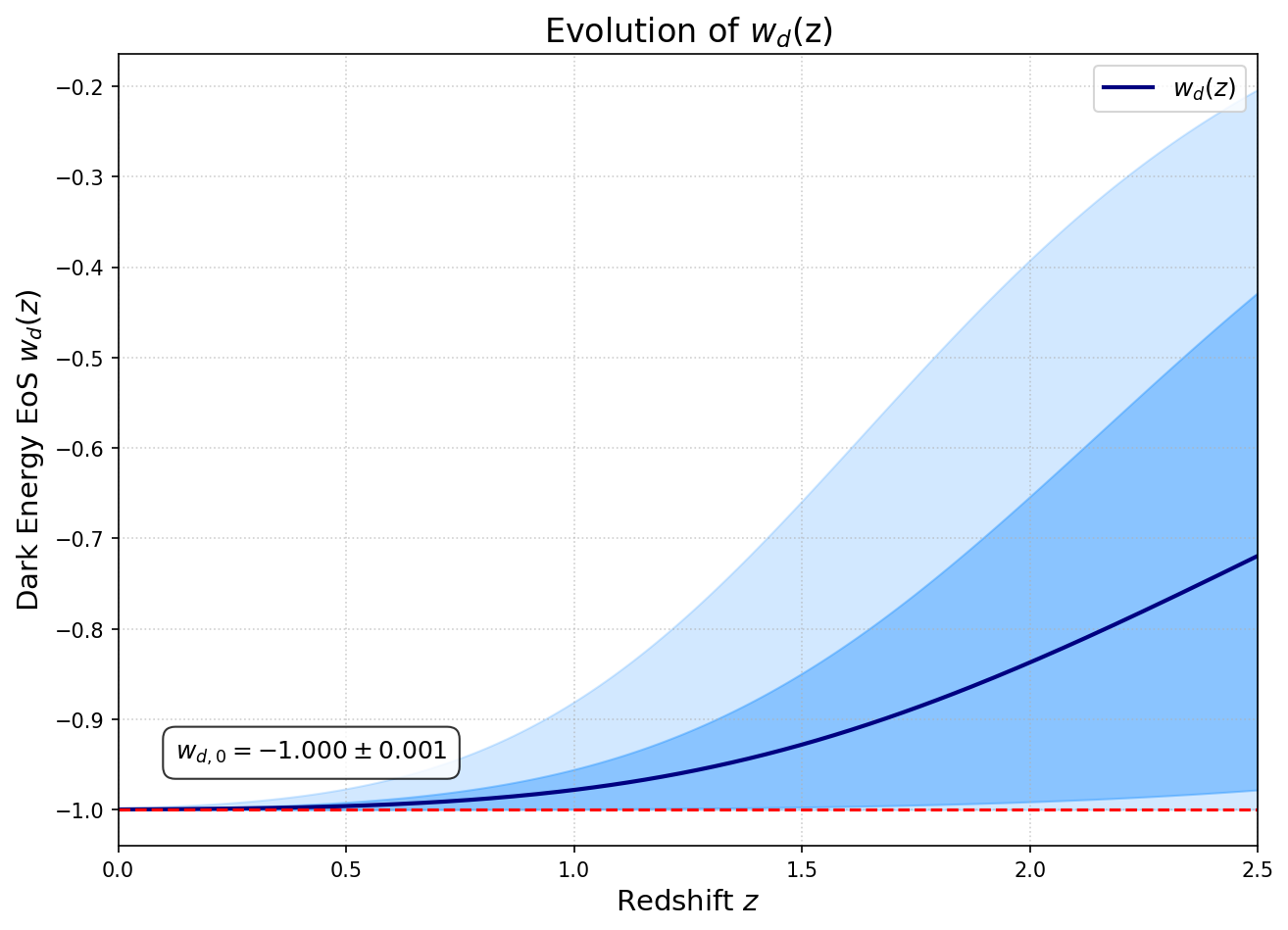}
    \caption{Plot of dark energy EoS parameter ($w_{d}$) with redshift distance ($z$) for Model II with 1$\sigma$ and 2$\sigma$ confidence interval. The $w=-1$ line corresponding to the $\Lambda$CDM model is shown as a red `dashed' line.}
    \label{Fig5} 
\end{figure}

\begin{figure}[ht] 
    \centering
    \includegraphics[width=\columnwidth]{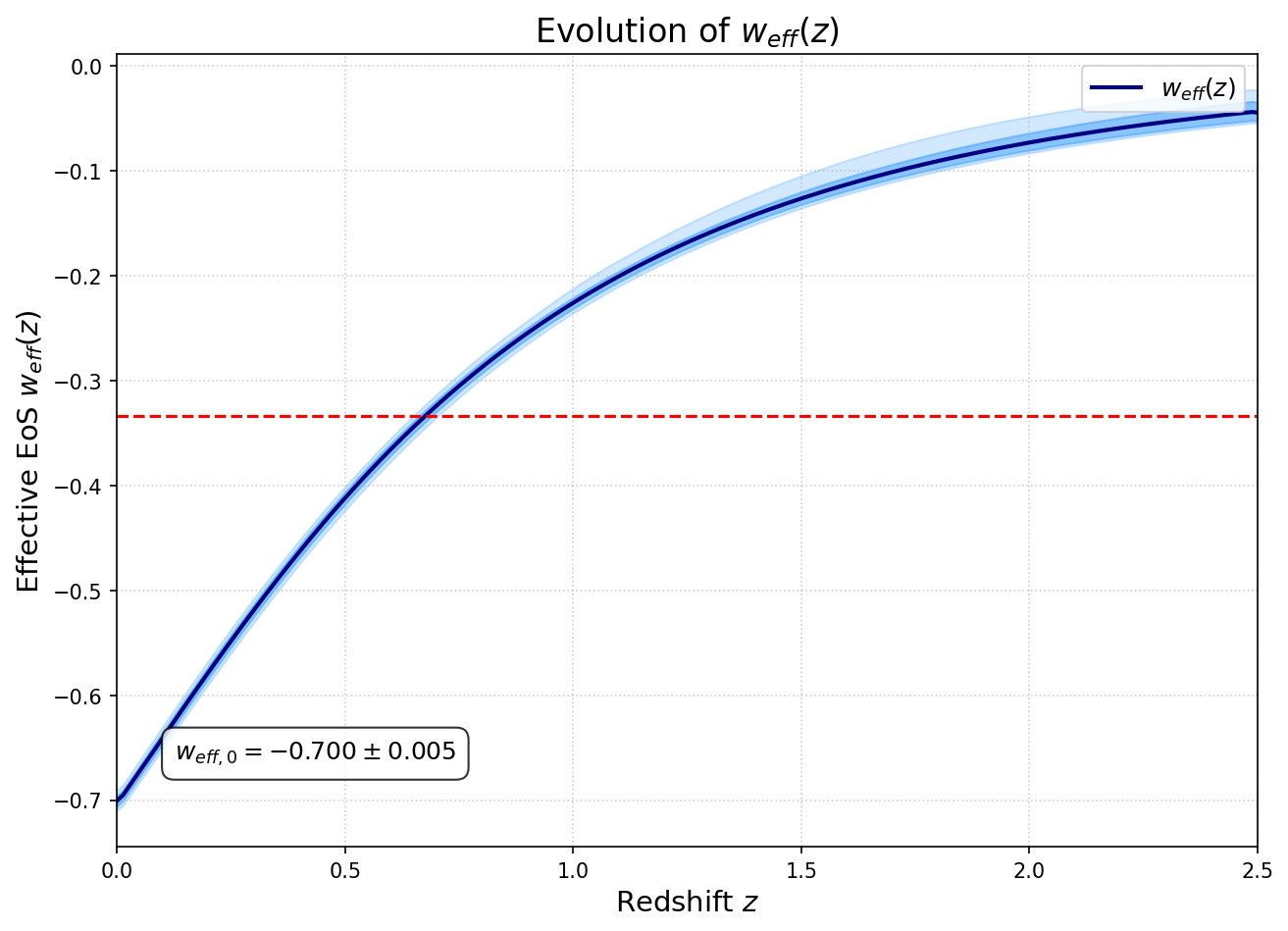}
    \caption{Plot of effective EoS parameter ($w_{eff}$) with redshift distance ($z$) for Model II with 1$\sigma$ and 2$\sigma$ confidence interval.}
    \label{Fig6} 
\end{figure}

We also plot the equation of state (EoS) parameter ($w_{\phi}\equiv w_{d}$ and $w_{\rm eff}$) [Eqs. (\ref{7}), (\ref{14}) and (\ref{24a})] with redshift distance ($z$) for both the model in Figs.~(\ref{Fig3}), (\ref{Fig4}), (\ref{Fig5}) and  (\ref{Fig6}) based on Bayesian inference using the ML-based surrogate method and reconstruct the functional dependence using the symbolic regression technique within the PySR framework~\cite{Meier} and the mean value of the derived EoS parameter. PySR framework combines evolutionary algorithms and high-performance computing. This yielded a compact analytical form for dark energy EoS from Fig. (\ref{Fig3}) as:  

\ben
w_{d}(z) \approx && 0.0772 z^2 + e^{z} \left(-0.0200\right) - 0.978 \n \\
\label{37}
\een

The evolution of the dark energy EoS, $w_{d}(z)$, presented in the Fig. (\ref{Fig3}), reveals several key features of the model.  At $z=0$, the dark energy EoS is found to be $w_{d,0} \simeq -0.997 \pm 0.007 $, remaining nearly constant during epochs where dark energy significantly influences the dynamics. It indicates that the DBI scalar field currently behaves similarly to a slowly varying vacuum energy component, closely resembling the behavior of cosmological constant. However, unlike $\Lambda$CDM, $w_{d}(z)$ exhibits a mild redshift dependence, gradually increasing toward less negative values at higher $z$. This behavior originates from the non-canonical kinetic structure of the DBI action, where both the warp factor $f(\phi)$ and the potential $V(\phi)$ shape the scalar field evolution and modify its effective pressure to density ratio over time. However, the widening of the uncertainty bands in $w_{d}(z)$ at $z \gtrsim 1$ reflects the decrease in constraining power of current cosmological observations at earlier epochs.

We also plot the effective equation of state parameter ($\o_{eff}$) for this model (Model I, Eq. \ref{8}) using Eq. (\ref{24a}). The plot of the effective EoS ($w_{eff}(z)$) is shown in the Fig. (\ref{Fig4}). Here, $w_{eff}(z)$ shows a rising behavior from negative values at low redshift to nearly zero at higher redshift, indicating a transition from a dark energy dominated era to a matter dominated era. At the present epoch ($z=0$), the Bayesian method yields $w_{eff}= -0.688 \pm 0.012$ , confirming an accelerating universe ($\o_{eff} < -\frac{1}{3}$ )\cite{Peebles1}.

Thus, Figs.~(\ref{Fig3}) and~(\ref{Fig4}) present a comparative study of the dark and effective energy EoS, $w_{d}(z)$ and $w_{eff}(z)$, respectively, inferred from Bayesian approach for Model I. Therefore, we may conclude that, this DBI  based scalar field model~(\ref{8}) stands as a compelling extension of $\Lambda$CDM. It retains all the successful phenomenology of the standard model at low redshift, but opens the door to richer dynamics at higher redshifts.\\

For Model II (\ref{9}), the analytic expression that we found from Fig. (\ref{Fig5}) for the dark energy equation of state parameter using the PySR technique~\cite{Meier} is expressed as:

\ben
w_{d}(z)\approx - e^{z^3 e^{z \left(-0.0515\right)} \left(-0.0243\right)}  
\label{38}
\een

The evolution of the effective and dark energy EoS parameters for the tachyonic DBI model (\ref{9}) or (\ref{14a}) is illustrated in Fig.~\ref{Fig5} and Fig.~\ref{Fig6}. 
The effective EoS parameter $w_{\rm eff}(z)$, shown in Fig.~\ref{Fig6}, displays a smooth transition from its present day value $w_{\rm eff,0} \simeq -0.700 $ towards less negative values at higher redshifts. 
This behavior reflects the gradual dilution of dark energy and the growing dominance of the matter component as we move back in cosmic time. The narrow shaded regions correspond to the $1\sigma$ uncertainty bands, indicating that the effective EoS is tightly constrained by the data across the entire redshift range.

Figure~\ref{Fig5} shows the reconstructed evolution of the dark energy EoS parameter $w_d(z)$ for the tachyonic DBI model (\ref{9}). At the present epoch, the model suggest present day dark energy EoS parameter value to be  $w_{d,0} \simeq -1.000$ and an uncertainty of $\sim 0.001$, indicating a near prefect $\L$CDM like behavior.  As we go to the higher redshift value, the equation of state parameter moves towards less negative value with greater uncertainties. The widening of the uncertainty bands at higher redshifts reflects the fact that cosmological observations are less sensitive to the behaviour of dark energy in the distant past, which allows for a larger range of model realizations.  Overall, both figures (Figs. \ref{Fig5} and \ref{Fig6}) suggest an dynamical dark energy behavior of the tachyonic DBI model that is completely different from $\L$CDM model at late time. 

A further theoretical consideration, not explored here, concerns the tracker like dynamics typical of k-essence scenarios. In these models, the scalar field adapts to the dominant background component, closely following the matter driven expansion before evolving into a dark energy dominated phase. The slight shift of the reconstructed EoS parameter toward less negative values may hint at such behaviour, since k-essence fields are expected to emulate the background fluid at intermediate redshifts and then deviate toward accelerated expansion at late times. Nonetheless, the restricted redshift range of our data prevents us from establishing the onset or duration of this tracking phase, and a targeted high redshift observations will be necessary to test this interpretation.

\section{Discussion of Results}
In this section, we discussed a statistical comparison to assess which model best fits the observational data. We benchmarked the results to show how significant our observational analysis via our proposed models is compared to the standard $\Lambda$CDM model ({\it viz. Appendix \ref{C}}) and $wCDM$ model with CPL parametrization ({\it viz. Appendix \ref{D}}). Additionally, we compute the deceleration parameter ($q$) for both models (Models I and II) to analyze their cosmological behaviors. Based on our observational data analysis, we construct a summarized table containing statistical criteria as follows:
\begin{table*}[ht]
\begin{center}
\resizebox{\textwidth}{2.0 cm}{%
    \begin{tabular}{|l|c|c|c|c|c|c|c|c|c|}
    \hline
    \textbf{Model} & \textbf{Method} & $ \chi^2 $ & $ \chi^2_{\nu} $ & AIC & BIC & DIC  & $\Delta$AIC  & $\Delta$BIC & $\D$DIC \\
    \hline
    $\L$CDM & (SVI)       & 1479.22 & 0.90 & 1485.22 & 1501.43 & 1485.20 & 0 & 0 & 0 \\
    \hline
    $w$CDM & (SVI) & 1471.25 & 0.90 & 1481.25 & 1508.26 & 1480.66 & - 3.97 & 6.83 & - 4.54 \\
    \hline
    Model I & (SVI)  & 1477.38  & 0.90  & 1489.38  & 1521.80 & 1489.14 & 4.16 & 19.63 & 3.94 \\
    \hline
    Model II & (SVI)  & 1479.15 & 0.91 & 1489.15 & 1516.16 & 1487.58 & 3.93 & 14.73 & 2.38 \\
    \hline
    \end{tabular}}
    \end{center}
    \caption{Comparison of statistical criteria for Model I and Model II  with $\L$CDM and $w$CDM using SVI  based Bayesian inference method}
    \label{TableIII}
    \end{table*}
here,
\begin{itemize}
    \item $ \chi^2 $: Total chi squared value.
    \item $ \chi^2_{\nu} $: Reduced chi squared, defined as \( \chi^2/\nu \), where $ \nu $ is the number of degrees of freedom, defined as \(\nu = N - k\), where $N=$ the number of data points and $k=$ number of model parameters.
    \item \textbf{AIC}: Akaike Information Criterion, defined as \( \mathrm{AIC} = \chi^2 + 2k \).
    \item \textbf{BIC}: Bayesian Information Criterion, defined as \( \mathrm{BIC} = \chi^2 + k \ln N \).
    \item \textbf{DIC}: Deviance Information Criterion, defined as $\mathrm{DIC} = \bar{D}(\theta) + p_D$, where $\bar{D}(\theta)$ is the posterior mean deviation and $p_D$ is the effective number of parameters
    \item \textbf{$\Delta$AIC}, \textbf{$\Delta$BIC} and \textbf{$\D$ DIC}: Difference between AIC, BIC and DIC for a particular model with respect to $\L$CDM model.
    
\end{itemize}

In addition to the aforementioned statistical measures, we also compute the \texttt{WAIC} and \texttt{PSIS-LOO} metrics based solely on Bayesian sampling. Since both of these criteria are inherently derived from posterior distributions, they are most appropriately evaluated using samples obtained from the Markov Chain Monte Carlo (MCMC) chains. The \texttt{WAIC} and the \texttt{PSIS-LOO} are fully Bayesian model comparison techniques that assess the predictive performance of models through the pointwise log-likelihood of the observed data. The WAIC is computed as  
\ben
\mathrm{WAIC} = -2 \left( \mathrm{LPPD} - P_{\mathrm{WAIC}} \right),
\label{39b}
\een
where the \texttt{log pointwise predictive density (LPPD)} is given by  
\ben
\mathrm{LPPD} = \sum_{i=1}^{N} \log \left( \frac{1}{S} \sum_{s=1}^{S} p(y_i \mid \theta^{(s)}) \right),
\label{39c}
\een
and the \texttt{effective number of parameters} is defined as  
\ben
P_{\mathrm{WAIC}} = \sum_{i=1}^{N} \mathrm{Var}_{s} \left[ \log p(y_i \mid \theta^{(s)}) \right],
\label{39d}
\een
where $ \theta^{(s)}$ denotes posterior samples from the model. 

The PSIS-LOO provides an alternative cross-validation-based estimate of model predictive performance by approximating exact leave-one-out cross-validation using importance sampling with Pareto-smoothed weights. The \texttt{expected log predictive density (ELPD$_{\text{loo}}$)} is calculated as  
\ben
\mathrm{ELPD}_{\text{loo}} = \sum_{i=1}^{N} \log \left( 
\frac{\sum_{s=1}^{S} w_i^{(s)} p(y_i \mid \theta^{(s)})}{\sum_{s=1}^{S} w_i^{(s)}} 
\right),
\label{39e}
\een
where $w_i^{(s)}$ are Pareto-smoothed importance weights. WAIC and PSIS-LOO both favor models that achieve superior out-of-sample predictive performance while simultaneously imposing a penalty on unnecessary model complexity. As a result, they serve as robust, Bayesianly well founded diagnostics for model comparison in cosmology, particularly when evaluating competing scenarios against combined datasets such as \PantheonShoes{}, BAO data, and cosmic chronometer data. To carry out the quantitative comparison between models in terms of WAIC and PSIS-LOO, we employ the \texttt{ArviZ} library within the Python ecosystem, which provides standardized, reliable tools for computing these information criteria from posterior samples.

In the table (\ref{TableIII}),  the number of data points ($N$) for all the models (Model I, Model II, $\Lambda$CDM and $w$CDM) is the same ($ N=1639$, data points for \PantheonShoes{} + Hubble+ BAO (DESIBAO- DR2)) while the number of model parameters ($k$) for Models I, Model II, $\L$CDM and $w$CDM  are respectively $k=6$, $k=5$, $k=3$ and $k=5$.

To evaluate and compare the performance of the two proposed models, Model I and Model II  against the benchmark $\L$CDM and $w$CDM model with CPL parametrization, we analyze their statistical metrics: $ \chi^2 $,  $ \chi^2_{\nu} $, AIC, BIC, DIC, WAIC and PSIS-LOO. These criteria collectively assess the goodness of fit and penalize model complexity, enabling a balanced model selection.

Table \ref{TableIII} summarizes the comparative statistical performance of the four cosmological scenarios $\Lambda$CDM, $w$CDM, Model I, and Model II evaluated within the SVI framework. The baseline $\Lambda$CDM model yields the smallest information criterion values and is therefore adopted as the reference, with $\Delta$AIC = $\Delta$BIC = $\Delta$DIC = 0.

The $w$CDM extension achieves a modest improvement in the deviance based measures, with reductions of $\Delta$AIC = $-3.97$ and $\Delta$DIC = $-4.54$ relative to $\Lambda$CDM. This points to a slightly better goodness of fit despite the introduction of one additional parameter. Nonetheless, the Bayesian Information Criterion penalizes this extra complexity more severely, yielding $\Delta$BIC = 6.83, which indicates that the improvement is not statistically compelling once model complexity is taken into account.

While the extended dynamical dark energy configurations (Model I and Model II) incur natural penalties for their additional complexity, they remain statistically competitive with the standard $\L$CDM and $w$CDM cosmologies. Although the information criteria (AIC, BIC, DIC) favor the simpler reference models with Model I showing the largest deviation ($\D \text{AIC} = 4.16$, $\D \text{BIC} = 19.63$, $\D \text{DIC} = 3.94$) these metrics primarily reflect a preference for parsimony rather than a failure of the physical model. Crucially, the $\D \text{AIC}$ and $\D \text{DIC}$ values relative to the $\L$CDM baseline do not exceed the threshold of $10$, meaning there is no decisive statistical evidence excluding these dynamical scenarios. Consequently, despite the strict penalties imposed by the SVI analysis on extra degrees of freedom, Models I and II provide a consistent fit to the current dataset and represent valid, physically richer alternatives to the standard paradigms.

Although the reduced chi square values for all four models remain close to unity, signalling statistically acceptable fits the information theoretic criteria, inspired by Occam’s razor \cite{Sober}, still favour the more parsimonious $\L$CDM model. Overall, these results demonstrate the statistical soundness of our DBI-type models within the adopted ML based Bayesian framework.

However, a key drawback of using \textit{AIC}, \textit{BIC}, and \textit{DIC} for model selection is that they depend only on the maximum log likelihood and penalize models with a larger number of free parameters. In cosmological applications, scenarios where dark energy evolves dynamically via a scalar field usually introduce extra parameters relative to simpler descriptions such as $\Lambda$CDM or $w$CDM. As a result, assessments based solely on AIC, BIC, and DIC are not fully reliable. To overcome this limitation, we employ more comprehensive Bayesian diagnostics, namely \texttt{WAIC} and \texttt{PSIS-LOO}. These criteria assess models using the pointwise log likelihood over all data points, instead of relying on a single maximum likelihood value. By making use of the full set of Bayesian posterior samples, they also infer an effective number of parameters directly from the data, rather than adopting the nominal parameter count specified by the user. This strategy yields a more dependable and statistically robust framework for comparing sophisticated cosmological models.

\begin{table*}[!t]
\centering
\renewcommand{\arraystretch}{0.9}
\footnotesize
\caption{Model comparison based on WAIC and PSIS-LOO}
\label{Table IV}
\resizebox{0.92\textwidth}{!}{
\begin{tabular}{|l|r|r|r|r|r|r|}
\toprule
Model & LOO (Deviance) & p\_LOO & d\_LOO & WAIC & p\_WAIC & Weight \\
\midrule
$w$CDM          & -1605.69 & 4.80 & 0.00 & -1608.42 & 3.43 & 0.25 \\
\hline
$\Lambda$CDM     & -1602.69 & 3.14 & 3.00 & -1603.06 & 2.96 & 0.25 \\
\hline
Model II      & -1600.80 & 3.66 & 4.88 & -1601.35 & 3.39 & 0.25 \\
\hline
Model I       & -1598.56 & 5.16 & 7.13 & -1600.12 & 4.38 & 0.25 \\
\bottomrule
\end{tabular}
}
\end{table*}

Table \ref{Table IV} shows that both DBI k-essence extensions, Model I and Model II, deliver predictive performance fully comparable to $\Lambda$CDM and $w$CDM when assessed using PSIS-LOO and WAIC cross-validation. While the LOO deviances of Model I ($-1598.56$) and Model II ($-1600.80$) are slightly higher than those of the simpler reference models, these shifts fall comfortably within the statistical noise of the PSIS reweighting procedure, leading to identical model weights ($0.25$) for all four scenarios. This degeneracy in weight allocation is significant, it implies that the observational data do not disfavour the extra dynamical degrees of freedom introduced by the DBI scalar sector, and that Model I and Model II remain on equal footing as viable descriptions of the late time expansion history. In addition, the effective parameter counts $p_{\mathrm{LOO}}$ and $p_{\mathrm{WAIC}}$ for the two DBI models stay relatively small, underscoring that their enriched k-essence dynamics do not induce problematic over flexibility or excessive model complexity. The somewhat elevated $d_{\mathrm{LOO}}$ values, especially for Model I ($7.13$), align with physically motivated deviations from a pure cosmological constant rather than with numerical overfitting, indicating that the DBI type kinetic structure can accommodate nontrivial dark energy evolution while preserving predictive robustness. Accordingly, the near equality of the cross-validation diagnostics with those of $\Lambda$CDM and $w$CDM reinforces the view that k-essence based DBI dynamics offer a credible extension of the standard cosmological framework, they reproduce the empirical success of $\Lambda$CDM yet maintain the theoretical flexibility needed to probe more fundamental questions regarding the origin, stability, and evolution of dark energy.

It is important to note that the Leave-One-Out (LOO) cross-validation calculations consistently flag a warning (\texttt{warning = True}) across all analyzed models, including the standard $\Lambda$CDM and $w$CDM baselines. This diagnostic arises naturally from the structure of the combined dataset rather than a specific model pathology. Specifically, the Pantheon+SHOES sample constitutes $1588$ of the $1639$ total data points. The inherent correlations within this dominant supernova dataset heavily influence the importance sampling weights, triggering the observed stability warning. Consequently, the Bayesian stacking analysis designed to optimize model combination for predictive accuracy assigns an approximately uniform weight distribution across all candidates. This implies that the current observational data do not possess sufficient constraining power to decisively prefer $\Lambda$CDM or $w$CDM over the DBI scalar field parametrizations (Model I and Model II). Crucially, this result contrasts sharply with a scenario of statistical exclusion, where alternative models would typically receive negligible weights (e.g., $< 0.01$). Instead, the comparable weights suggest that Model~I~(\ref{8}) and Model~II~(\ref{14a}) occupy likelihood regions that substantially overlap with the concordance cosmology. This reinforces the findings from the reduced $\chi^2$ analysis ($\chi^2_\nu \approx 0.90$), confirming that Models I and II remain observationally indistinguishable from the standard model and stand as statistically viable alternatives, distinguishable primarily by their theoretical motivation rather than empirical disadvantage.

\begin{figure}[ht] 
    \centering
    \includegraphics[width=\columnwidth]{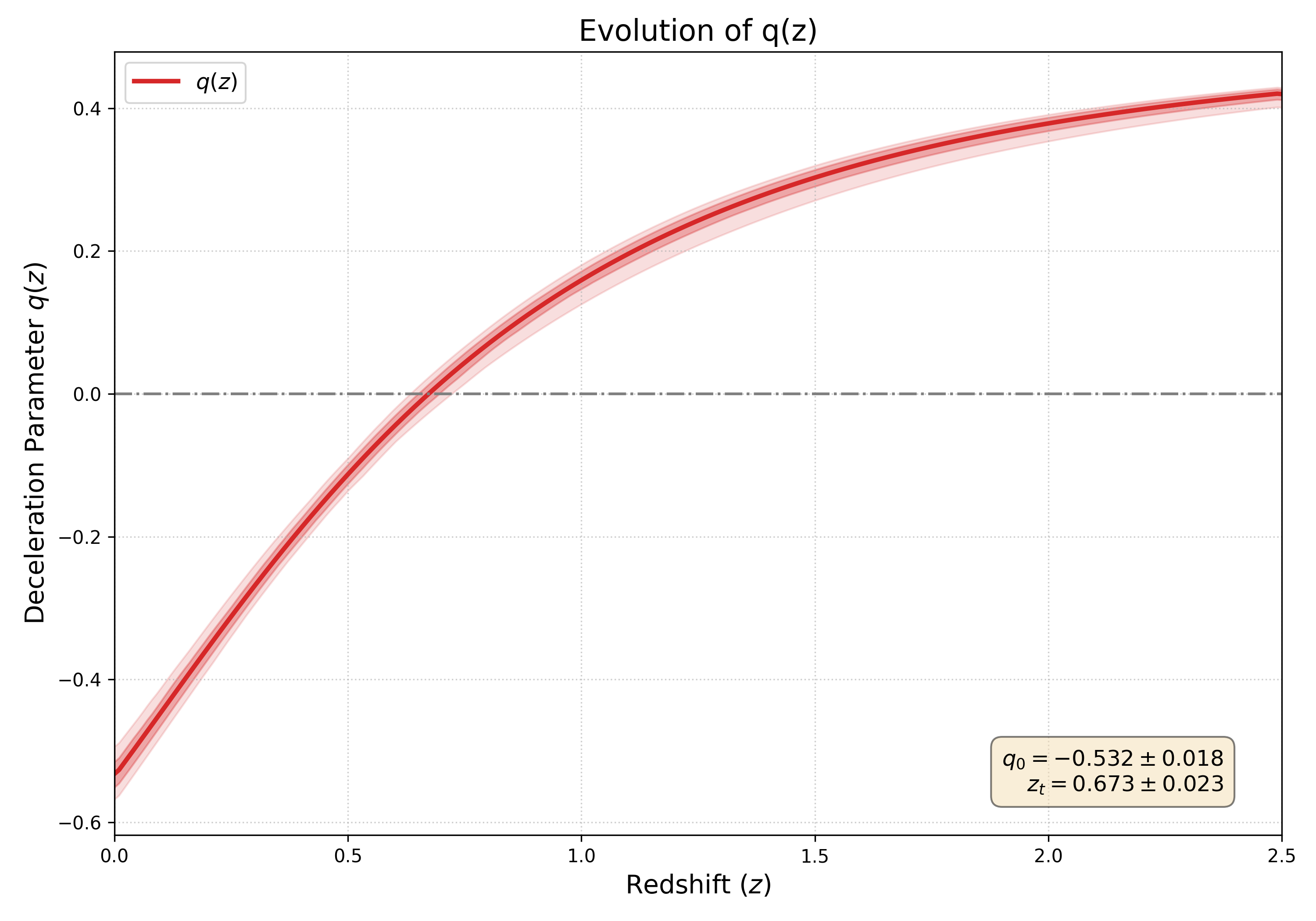}
    \caption{Plot of deceleration parameter~($q$) with redshift distance ($z$) for Model I}
    \label{Fig7} 
\end{figure}

\begin{figure}[ht] 
    \centering
    \includegraphics[width=\columnwidth]{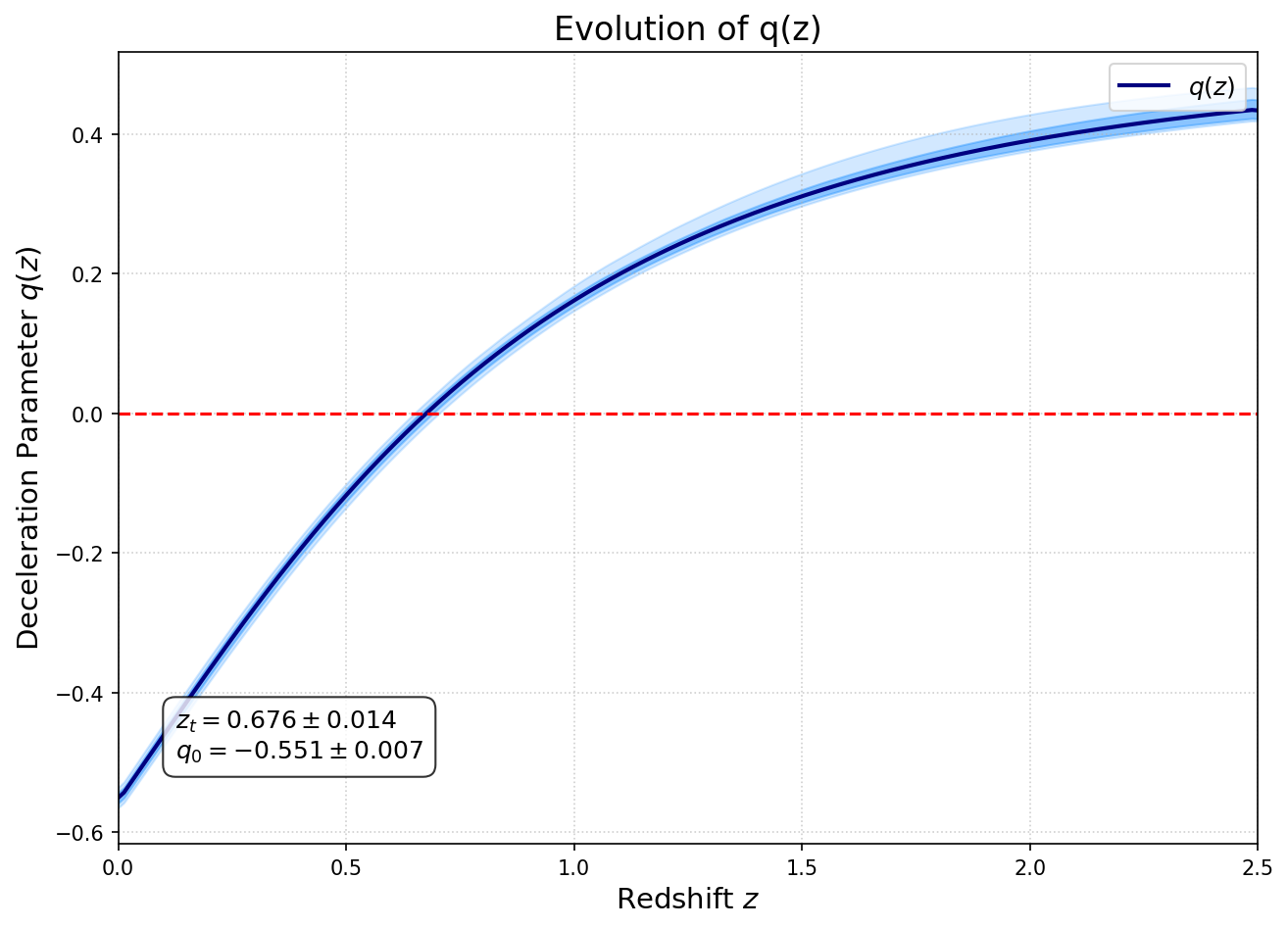}
    \caption{Plot of deceleration parameter~($q$) with redshift distance ($z$) for Model II}
    \label{Fig8} 
\end{figure}

Although Model II has slight statistical edge over Model I, a more comprehensive judgment requires not only quantitative model comparison but also a physical interpretation of the underlying dynamics. This is achieved by analyzing the behavior of the deceleration parameter~$\big(q= -1 - \frac{\frac{dH}{dt}}{H^2}= -1 + (1+z)\frac{\frac{dH}{dz}}{H}\big)$, which encapsulates the acceleration history of the universe. By estimating the model parameters through observational constraints and subsequently computing the redshift evolution of $q(z)$, we gain deeper insight into the cosmic expansion trends characteristic of each model. The resulting deceleration profiles, presented in Fig. (\ref{Fig7}) and Fig. (\ref{Fig8}), visually demonstrate the similarity in acceleration mechanisms.  

The comparative study of the two k-essence dynamical dark energy scenarios, Model I and Model II, shows that they exhibit cosmological behavior similar to that of the reference $\Lambda$CDM and $w$CDM models. These latter models continue to serve as the standard framework, supported by high precision observations including Planck CMB data \cite{Planck5}, BAO measurements \cite{Desi}, and Type Ia supernovae \cite{Scolnic1}.

Model I, constructed from a DBI like k-essence Lagrangian of the form $\La=-f(\phi)\sqrt{1- \frac{2X}{f(\phi)}}+ f(\phi) - V(\phi)+\La_{m}$ with $f(\phi)=\l \phi^4$ and $V(\phi)= \frac{m^2\phi^2}{2}$, predicts a present day deceleration parameter $q_{0}^{Model-I}= - 0.532 \pm 0.018 $ and a transition redshift $z_{trans}=0.673 \pm 0.023$. These values indicate that cosmic acceleration is relatively weaker for DBI type dark energy model while transition redshift is comparable to both $\L$CDM and $w$CDM model, which typically gives $q^{\L CDM}(0)= - 0.551 \pm 0.010 $  with $z_{trans} = 0.672 \pm 0.017$ ({\it Appendix} Fig. \ref{Fig10}) \cite{Pareek, Riess1} and $q^{w CDM}(0)= - 0.402 \pm 0.055 $  with $z_{trans} = 0.708 \pm 0.028$ ({\it Appendix} Fig. \ref{Fig12}).

By contrast, Model II, with its simplified DBI-type Lagrangian 
$\La(\phi, X)=-V(\phi)\sqrt{1-2X}$ and the same potential $V(\phi)= \frac{m^2\phi^2}{2}$, yields a significantly stronger acceleration today $q^{Model-II}_{0} = - 0.551 \pm 0.007$, with a transition at $z_{trans}= 0.676 \pm 0.014$. The present-day deceleration value and transition redshift for Model II are in excellent agreement with those of standard $\L$CDM model. 

From a physical perspective, the choice between these models hinges on which one is considered more fundamental. Within our adaptation of $\Lambda$CDM as the reference framework, both Model I and Model II appear as well-motivated extensions of the dynamical dark energy model. Each maintains a $\Lambda$CDM like evolution of the deceleration parameter, produces a moderately increased value of $H_0$ consistent with local measurements, and restricts the variations in $\Omega_{m0}$ and $r_d$ to be small enough not to substantially degrade the early-time fits. Their dynamical behavior is compatible with slowly varying scalar field descriptions and involves only mild departures from the standard cosmological scenario.

Although our result shows profound consistencies with the benchmark model, we acknowledge that a complete assessment of any cosmological model requires a joint analysis with early universe probes, most notably the CMB data from Planck \cite{Planck5}. However, CMB analysis is beyond the scope of this study.

\section{Conclusion}
We may conclude that both of our models (Models I and II) can reproduce key features of cosmic expansion and are compatible with current observational datasets (Pantheon+SHOESSN Ia, Hubble, and BAO) for the late universe. Therefore, the key inferences are as follows: 
\begin{itemize}
    \item In this analysis, the DBI framework provides a physically motivated scalar field description that behaves as an effective fluid and possesses a well-defined dynamical equation of motion, capable of adjusting the dark energy evolution of the universe. In this sense, it provides a theoretically grounded alternative to the cosmological constant or purely phenomenological parameterizations of the dark energy equation of state, offering a coherent theoretical basis for interpreting late-time cosmic acceleration. We therefore place our model within the broader class of alternative gravity theories, where the scalar field dynamics are intrinsically linked to the observed expansion history of the universe.
    
    \item Our analysis establishes a decisive methodological advancement in cosmological parameter inference. By replacing conventional MCMC sampling with a ML accelerated surrogate model integrated with SVI, we achieve full Bayesian posterior reconstruction with orders of magnitude improvement in efficiency. Model selection is executed using fully Bayesian statistical diagnostics specifically WAIC and PSIS-LOO, which provide robust, information theoretic discrimination across competing dynamical dark energy scenarios. These results confirm that ML driven surrogate inference, combined with principled Bayesian model comparison, constitutes a superior and scalable alternative to traditional sampling dependent methodologies.
    
    \item Our analysis identifies Model II (\ref{14a}) as the statistically and physically preferred model. It provides a slightly better fit to the data, as demonstrated by lower values of $\chi^2$, AIC, BIC and DIC, and offers a more $\Lambda$CDM-like evolution of the deceleration parameter. It also admits a Hubble constant $H_0$ comparable to $\Lambda$CDM and $w$CDM (Table~\ref{Table-IVa}), consistent with local measurements when analyzed with the late-time datasets considered here. Regarding the inferred sound horizon, our values $r_d \simeq 138$--$139\,\mathrm{Mpc}$ from both models are in close agreement with the late-universe BAO analysis of Liu et al.~\cite{Liu}, who obtain a consistent range without CMB priors. The $\sim 9\,\mathrm{Mpc}$ offset from the Planck 2018 value $r_d = 147.09 \pm 0.26\,\mathrm{Mpc}$~\cite{Planck5} is not a model pathology but a direct consequence of the well-known $H_0$--$r_d$ degeneracy. Since BAO data primarily constrain the combination $H_0 r_d / c \approx \mathrm{const.}$, anchoring $H_0 \sim 73\,\mathrm{km\,s^{-1}\,Mpc^{-1}}$ from Pantheon$+$SH0ES drives the posterior toward a lower effective $r_d$, regardless of the dark energy model. As the sound horizon $r_d$ is an \emph{early-universe} scale set by pre-recombination physics (e.g.\ the baryon and photon densities and the radiation content) and therefore cannot be robustly reconstructed from late-time data alone without a CMB prior. Our inferred $r_d$ should thus be interpreted as an \emph{effective} BAO distance scale within a late-universe fit, rather than as a Planck-consistent early-universe sound horizon. However, recovering a Planck-consistent $r_d$ would require incorporating CMB information into the likelihood, which is beyond the scope of the present work.

    \item Although the DBI models do not strictly surpass the benchmark $\Lambda$CDM or $w$CDM scenarios (with CPL parametrization) under the traditional information criteria (AIC, BIC, DIC), a comprehensive evaluation using fully Bayesian model comparison techniques specifically WAIC and PSIS-LOO provides a more favorable perspective. Within this rigorous framework, which prioritizes point-wise predictive accuracy, we find that both of our proposed Model I and Model II exhibit performance levels comparable to the reference cosmologies. Consequently, despite the additional parametric complexity, these models emerge as statistically tenable alternatives, demonstrating sufficient Bayesian support to be regarded as competitive descriptions of the dark energy sector alongside the standard concordance model.

\end{itemize}

\begin{table*}[ht]
\begin{center}
\resizebox{0.8\textwidth}{1.5 cm}{%
    \begin{tabular}{|c|c|c|c|}
    \hline
    Model & $H_0~[Km/s/Mpc]$ & $\O_{m0}$ & $r_d~[Mpc]$ \\
    \hline
    $\L$CDM & $73.56 \pm 0.15$  & $0.299 \pm 0.007$ & $137.82 \pm 0.67$ \\
    \hline
    $w$CDM & $72.65 \pm 0.27$ & $0.3073 \pm 0.0099$ & $136.95^{+0.70}_{-0.84}$ \\
    \hline
    Model I & $ 73.35 \pm 0.16$ & $0.307 \pm 0.011$ & $138.08 \pm 0.71$ \\
    \hline
    Model II & $73.35 \pm 0.15$ & $0.2960 \pm 0.0049$ & $138.81 \pm 0.57$ \\
    \hline
    \end{tabular}}
    \caption{Comparative Table of Cosmological Parameters for Different Models}
    \label{Table-IVa}
\end{center}
\end{table*}

The outcome based on statistical analysis is especially noteworthy because, according to the BIC criterion, our models are strongly disfavored as the values of $\Delta$BIC exceed 10, which is usually interpreted as decisive evidence against models with higher complexity. In contrast, the WAIC and PSIS-LOO results demonstrate that, from the standpoint of predictive accuracy and goodness of fit, Model I and Model II perform comparably to the standard benchmark models. Thus, while BIC penalizes the additional parameters in our extensions, the fully Bayesian predictive criteria suggest that these more complex models remain competitive in terms of their ability to fit and generalize to new data.
    
In one of our recent studies \cite{Ganguly}, we found that using a non affine parametrization of the Raychaudhuri equation in a purely kinetic DBI type k-essence framework ($\La(X)=-\sqrt{1-2X}$) led to a deceleration parameter of $q=-0.58$ and a transition redshift of $z_{trans}=0.73$. These values imply that the universe began accelerating earlier in its history and that dark energy had a significant influence even at higher redshifts. This points to a scenario where dark energy domination sets in earlier, possibly modifying structure formation and affecting the observable signatures in high redshift datasets.

In contrast to the analysis presented in \cite{Ganguly}, our current study employs the same kinetic structure but introduces an additional potential term, $V(\phi)$, in Model II. This modification yields $q_{0} = -0.551 \pm 0.007$, with the corresponding transition redshifts $z_{\mathrm{trans}} = 0.676 \pm 0.014$. The inferred transition redshift lies closer to the $\Lambda$CDM prediction, suggesting that the inclusion of the potential term brings the expansion dynamics of the DBI model into closer agreement with the standard cosmological model. This further implies that the non-affine parameter considered in \cite{Ganguly}$\,$ may have induced an enhanced acceleration in the expansion history, thereby yielding a comparatively larger magnitude of the deceleration parameter than obtained in the present analysis.

A persistent challenge confronting every alternative theory of gravity and dark energy is the so called \textit{Hubble tension}, the discrepancy between the locally measured value of the Hubble constant, $H_0$, from distance ladder calibrations~(SHOES) and the lower value inferred from the cosmic microwave background (CMB) under the standard $\Lambda$CDM framework. This tension reflects a deeper degeneracy between the present day expansion rate, $H_0$, and the sound horizon scale at baryon drag, $r_d$, which jointly determine the observed distance measures across the low and high redshift universe. In our investigation of two DBI based k-essence dark energy models, we find that late time observables such as BAO, Supernova (Pantheon + SH0ES), and cosmic chronometer data are insufficient to break this degeneracy on their own. Although these datasets can reproduce the observed expansion history across varying combinations of $H_0$ and $r_d$, they lack sensitivity to the early-universe physics that fundamentally fixes $r_d$. A thorough investigation of the CMB is needed to determine if DBI-type dynamics can influence the sound horizon by altering the pre-recombination expansion history, which may contributing to the resolution of the Hubble tension within a unified dynamical dark energy model. However, the analysis of the CMB is beyond the scope of the current investigation. The current study focuses solely on the late-time universe, while a comprehensive treatment from the early to the late universe will be pursued in future studies.

In summary, from a methodological standpoint, this work demonstrates the complementary power of Bayesian sampling with ML based surrogate model in cosmological model inference. The Bayesian framework captures full posterior distributions and parameter degeneracies, while ML accelerates likelihood evaluations, enabling rapid exploration of complex models. Moreover, by adopting WAIC and PSIS-LOO, we replace maximum log likelihood driven, parameter dependent criteria such as AIC and BIC with fully Bayesian model comparison. Although our results still identify $w$CDM as the preferred model, but considerations of model complexity and parsimony do not exclude our DBI extended dark energy scenario within a full Bayesian framework. Overall, our findings underscore the potential of DBI type k-essence models with standard action, particularly the unified variant (Model II (Eq. (\ref{14a})), as viable dynamical alternatives to $\Lambda$CDM or $w$CDM. Although many other DBI-type k-essence scenarios are conceivable, they lie outside the present scope of this work. Future work incorporating structure formation, non-linear perturbations, and forthcoming data (e.g., Euclid, LSST, JWST) will be essential to further test Models I and II and to sharpen constraints on the dark sector.\\

{\bf Acknowledgement:}
The authors thank Prof. Surhud More, IUCAA, Pune, for his valuable and illuminating suggestions on the manuscript. G.M. (Visiting Associate, IUCAA) and S.G. gratefully acknowledge IUCAA, Pune, India, for the visiting opportunity and for facilitating the completion of part of this work during their visits. A part of the work of A.B. has been supported by RUSA project of the University of Calcutta. E.G. and G.M. also thank the COST Association (CA21136 CosmoVerse), European Union, for the opportunity to participate in this international association as group members. G.M. extends heartfelt thanks to all the undergraduate, postgraduate, and doctoral students, as well as to all the teachers, collaborators, and well-wishers, whose support has significantly enriched him.  S.G. gratefully acknowledges his father for his unwavering support in shaping his character and guiding his personal growth, and extends sincere appreciation to all co-authors for their valuable and insightful contributions.
\\

{\bf Conflicts of interest:} The authors declare no conflicts of interest.\\

{\bf Funding information:} Not available.\\

{\bf Data availability:} The data used in this study are readily accessible from public sources for validation of our model; however, we did not generate any new datasets for this research.\\

{\bf Declaration of competing interest:}
The authors declare that they have no known competing financial interests or personal relationships that could have appeared to influence the work reported in this paper.\\

{\bf Declaration of generative AI in scientific writing:} The authors state that they do not support the use of AI tools to analyze and extract insights from data as part of the study process.\\

\appendix
\section{Hubble and BAO data set}
\label{A}

The Hubble parameter measurements $H(z)$ presented in Table~\ref{TableIV} are compiled from a range of observational studies spanning redshifts from $z\sim 0.7$ to $ z\sim 1.965$. These data are derived using the cosmic chronometer (or differential age) method, which estimates $H(z)$ by measuring the age difference of passively evolving galaxies at different redshifts.

\begin{table}[ht]
    \centering
    \begin{tabular}{|c|c|c|c|c|}
        \hline
        z & H(z) & $\sigma_H$ & Ref. \\
        \hline
        0.07 & 69.0 & $\pm19.68$ & \cite{Zhang} \\
        0.09 & 69.0 & $\pm 12.0$ & \cite{Simon} \\
        0.12 & 68.6 & $\pm 26.2$ & \cite{Zhang} \\
        0.17 & 83.0 & $\pm 8.0$  & \cite{Simon} \\
        0.179 & 75.0 & $\pm 4.0$ & \cite{Moresco1}  \\
        0.199 & 75.0 & $\pm 5.0$ &\cite{Moresco1} \\
        0.2 & 72.9 & $\pm29.6$   & \cite{Zhang} \\ 
        0.27 & 77.0 & $\pm14.0$  & \cite{Simon} \\
        0.28 & 88.8 & $\pm 36.6$ & \cite{Zhang} \\
        0.352 & 83.0 & $\pm 14.0$ & \cite{Moresco1} \\
        0.38 & 83.0 & $\pm 13.5$ & \cite{Moresco2} \\
        0.4 & 95 & $\pm 17.0$ & \cite{Simon} \\
        0.4004 & 77.0 & $\pm 10.2 $ & \cite{Moresco2} \\
        0.425 & 87.1 & $\pm 11.2$ & \cite{Moresco2} \\
        0.445 & 92.8 & $\pm12.9$ & \cite{Moresco2} \\
        0.47 & 89 & $\pm50$ & \cite{Ratsimbazafy} \\
        0.4783 & 80.9 & $\pm 9.0$ & \cite{Moresco2} \\
        0.48 & 97.0 & $\pm 62.0$ & \cite{Stern} \\
        0.593 & 104.0 & $\pm13.0$ & \cite{Moresco1} \\
        0.68 & 92.0 & $\pm8.0$ & \cite{Moresco1} \\
        0.75 & 98.8 & $\pm 33.6$ & \cite{Borghi} \\
        0.781 & 105.0 & $\pm12.0$ & \cite{Moresco1} \\
        0.875 & 125.0 & $\pm17.0$ &\cite{Moresco1} \\
        0.88 & 90.0 & $\pm40.0$ &\cite{Stern} \\
        0.9 & 117.0 & $\pm23.0$ & \cite{Simon} \\
        1.037 & 154.0 &$\pm20.0$ & \cite{Moresco1} \\
        1.3 & 168.0 & $\pm17.0$ & \cite{Simon} \\
        1.363 & 160.0 & $\pm33.6$ & \cite{Moresco3} \\
        1.43 & 177.0 & $\pm18.0$ & \cite{Simon} \\
        1.53 & 140.0 & $\pm14.0$ & \cite{Simon} \\
        1.75 & 202.0 & $\pm40.0$ & \cite{Simon} \\
        1.965 & 186.5 & $\pm50.4$ & \cite{Moresco3} \\
        \hline
    \end{tabular}
    \caption{Here the unit of $H(z)$ is $km s^{-1} Mpc^{-1}$,the $H(z)$ value is deduced from cosmic chronological method/differential age method and the corresponding reference from where the data are collected is mentioned in column Ref}
    \label{TableIV}
\end{table}

In this work, we perform a detailed analysis of baryon acoustic oscillation (BAO) measurements. Specifically, we focus on a set of seven BAO data points obtained by the Dark Energy Spectroscopic Instrument (DESI) collaboration \cite{Desi}, which we collectively refer to as DESIBAO. For our investigation, we employ the most recently available data release from DESI (DR2), ensuring that our results are based on the latest observational constraints. Throughout this study, we designate this DESIBAO compilation as our BAO dataset. A complete summary of the individual measurement, together with their associated redshifts and uncertainties, is presented in Table \ref{Table3}, where the full structure and content of the dataset are documented in detail.


\begin{table*}[ht]
    \centering
    \renewcommand{\arraystretch}{1.4}
    \resizebox{\textwidth}{!}{%
    \begin{tabular}{|c|c|c|c|c|c|c|c|}
        \toprule
        $\t{z}$ & \textbf{0.295} & \textbf{0.510} & \textbf{0.706} & \textbf{0.934} & \textbf{1.321} & \textbf{1.484} & \textbf{2.330} \\ \midrule
        $D_V(\tilde{z})/r_d$ & 7.942 $\pm$ 0.075 & 12.720 $\pm$ 0.099 & 16.050 $\pm$ 0.110 & 19.721 $\pm$ 0.091 & 24.252 $\pm$ 0.174 & 26.055 $\pm$ 0.398 & 31.267 $\pm$ 0.256 \\ \midrule
        $D_M(\tilde{z})/r_d$ & & 13.588 $\pm$ 0.167 & 17.351 $\pm$ 0.33 & 21.576 $\pm$ 0.152 & 27.601 $\pm$ 0.318 & 30.512 $\pm$ 0.760  & 38.988 $\pm$ 0.531 \\ \midrule
        $D_H(\tilde{z})/r_d$ & & 21.863 $\pm$ 0.425 & 19.455 $\pm$ 0.330 & 17.641 $\pm$ 0.193 & 14.176 $\pm$ 0.221 & 12.817 $\pm$ 0.516 &  8.632 $\pm$ 0.101   \\
        \bottomrule
    \end{tabular}}
    \caption{DESIBAO measurements from DR2}
    \label{Table3}
\end{table*}

\section{Choice of parameter for Model I and Model II}
\label{B}
We select a six parameters ({\bf p}=($H_0$, $\Omega_{m0}$, $\phi_0$, $x_0$, $r_d$, $\l$)) differential equation for Model I while restricting the parameters to five ({\bf p}=($H_0$, $\Omega_{m0}$, $r_d$, $\phi_0$, $v_0$)) for Model II. Our choice is solely based on the model’s complexity and the redundancy of additional parameters in observational model fitting.

In constructing the first model, we exclude $m$ as an independent parameter by exploiting the constraint satisfied by the scalar field, namely

\ben
\Omega_{d0}= \frac{(\gamma_0 -1)\lambda \phi_0^4 + \dfrac{m^2 \phi_0^2}{2}}{3 H_0^2}\,,
\label{B1}
\een

from which we can solve for $m$ as

\ben
m = \sqrt{\frac{2\left(3 H_0^2 \Omega_{d0} - (\gamma_0 -1)\lambda \phi_0^4\right)}{\phi_0^2}}\,.
\label{B2}
\een

Consequently, once the parameters \(H_0\), \(\lambda\), \(\phi_0\), \(v_0\), and \(\Omega_{d0} = 1 - \Omega_{m0}\) are specified, the mass parameter \(m\) is uniquely determined by Eq. \eqref{B2}. In other words, \(m\) does not constitute a truly free parameter of the model but is instead constrained by the chosen values of the other quantities. Hence, treating \(m\) as an additional independent parameter would introduce redundancy, and we therefore omit it from the set of fundamental model parameters.

For Model II, we likewise refrain from treating $m$ as a fundamental free parameter in the fitting procedure. In this case as well, we can employ the same relation as above,
\ben
\Omega_{d0}= \frac{m^2\phi_0^2\sqrt{1 - v_0^2}}{6 H_0^2}
\label{B3}
\een
from which, by solving for $m$, we obtain
\ben
m = \sqrt{\frac{6 H_0^2 \Omega_{d0}}{\phi_0^2\sqrt{1 - v_0^2}}}
\label{B4}
\een

This expression (\ref{B4}) shows that $m$ is not an independent quantity but is instead determined by the late time cosmological parameters $\Omega_{d0}= 1 - \O_{m0}$, $H_0$, $\phi_0$, and $v_0$. Consequently, if we were to treat $m$ as a separate free parameter in the statistical analysis, we would merely enlarge the dimensionality of the parameter space without obtaining any substantial improvement in the quality or flexibility of the model fits.

In other words, Eq. \eqref{B1} and \ref{B3} effectively acts as a constraint equation that we impose on the theory. By enforcing this relation, we ensure that the late time universe is composed solely of the total matter and dark energy components, with the dark energy density parameter $\Omega_{d}$ fixed consistently with the remaining cosmological parameters.

\section{$\L$CDM Model}
\label{C}

We consider the usual $\L$CDM model and perform a Bayesian analysis to compare it with two of our proposed models (Model I (\ref{8}) and Model II (\ref{14a})) under consideration against the combined dataset of Pantheon+SHOES, Hubble, and BAO. In this model ($\L$CDM), the energy content of the universe is assumed to be dominated by two components: pressureless total matter with energy density $\rho_m$, and dark energy represented by a cosmological constant $\L$, with energy density $\rho_d$.

Assuming a spatially flat FLRW background, the dynamics of the universe are governed by the Friedmann equations:
\ben
&&H^2 = \rho_m + \rho_d
\label{C1} \\
&&\frac{\Ddot{a}}{a} = -\frac{1}{6}\Big(\rho_m+ 3P_m + \rho_d + 3P_d \big)
\label{C2}
\een
The energy densities satisfy the continuity equation as follows:
\ben
&&\frac{d\rho_m}{dt} = -3 H\rho_m \n \\
&& \frac{d\rho_d}{dt} = 0.
\label{C3}
\een
We define ($\Omega_{\L}$) and  ($\Omega_m$) as $\O_{\L}= \frac{\rho_d}{3H^2}$, and $\O_m =\frac{\rho_m}{3 H^2}$ respectively. These lead to the differential equation for $H$ and $\O_m$ as,
\ben
&&\frac{dH}{dz} = \frac{3 H}{2(1+z)}\O_m
\label{C4} \\
&& \frac{d\Omega_m}{dz} = -\frac{2 \Omega_m}{H}\frac{dH}{dz} + \frac{3 \Omega_m}{(1 + z)}
\label{C5}
\een
Together with Eq. (\ref{C5}), the condition $\Omega_{m0} + \Omega_{\L 0} =1$ specifies the total energy content of a spatially flat universe. 

The set of Eqs.~(\ref{C4}), (\ref{C5}),  (\ref{17}) along with the condition $\Omega_{m0} +\Omega_{\L0} =1$, allows us to fully determine the evolution of the Hubble parameter and the energy content of the universe.

\begin{table*}[ht]
\centering
\begin{tabular}{|c|c|c|}
\hline
\textbf{Parameter} & \textbf{Prior} & \textbf{SVI best-fit} \\
\hline
$H_0$[km/s/Mpc] & $\mathcal{U}(60,80)$ & $73.56 \pm 0.15$ \\
\hline
$\Omega_{m 0}$ & $\mathcal{U}(0.2,0.4)$ & $0.299 \pm 0.007$  \\
\hline
$r_d$[Mpc] & $\mathcal{U}(120,170)$ & $137.82 \pm 0.67$  \\
\hline
\end{tabular}
\caption{Parameter estimates and uncertainties using Bayesian inference with NumPyro via SVI optimization  for the $\Lambda$CDM model.}
\label{TableVIII}
\end{table*}

\begin{figure}[H] 
    \centering
    \includegraphics[width=\columnwidth]{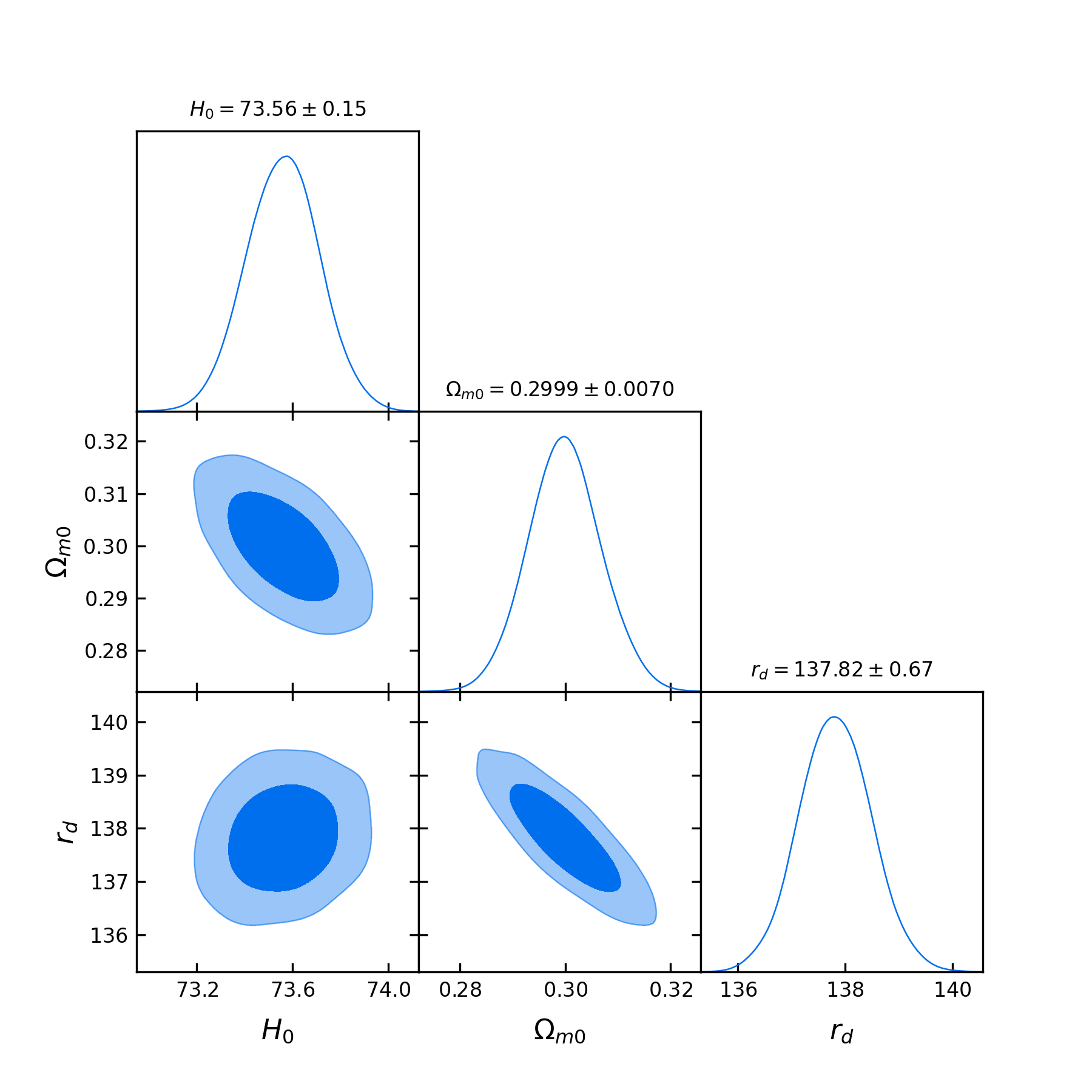}
    \caption{Bayesian analysis using SVI for the combined data set of Hubble+ Pantheon+SHOES BAO for $\L$CDM Model}
    \label{Fig9} 
\end{figure}

For completeness, we also plot the deceleration parameter $q$ as a function of $z$ for the $\Lambda$CDM model, showing the 1$\sigma$ and 2$\sigma$ confidence regions based on the best fit parameters obtained from the SVI Bayesian analysis. The transition redshift $z_t$ and the current value $q_0$ are indicated on the plot together with their standard deviations.

\begin{figure}[H] 
    \centering
    \includegraphics[width=\columnwidth]{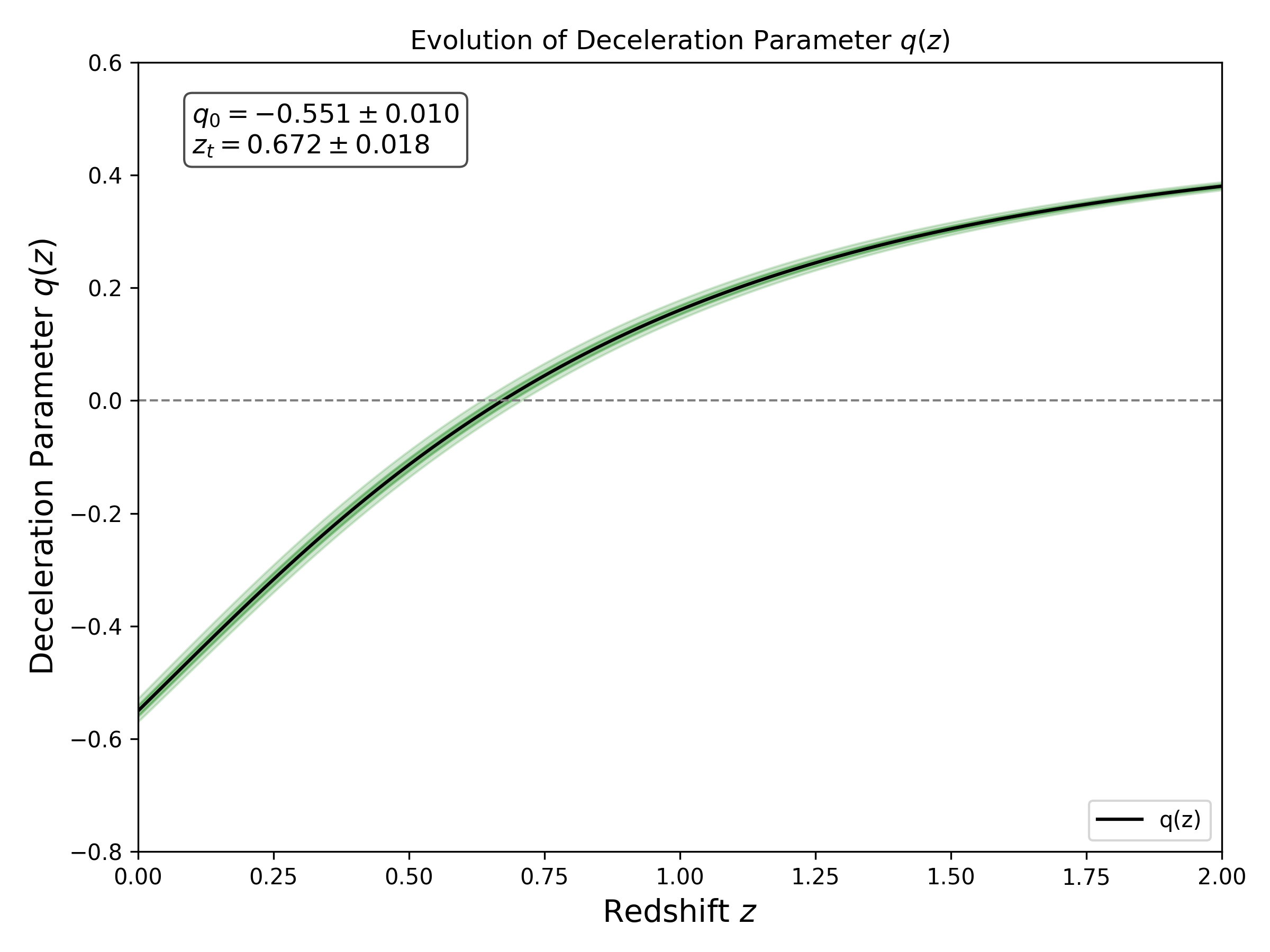}
    \caption{Plot of deceleration parameter~($q$) with redshift distance ($z$) for $\L$CDM Model}
    \label{Fig10} 
\end{figure}




\section{$w$CDM Model}
\label{D}
For comparison, we also carry out a Bayesian analysis using the combined Pantheon+SHOES, Hubble, and BAO datasets within the framework of the flat $\Lambda$CDM model extended by the Chevallier-Polarski-Linder (CPL) parametrization~\cite{Chevallier} of the dark energy equation of state (EoS). The CPL form is given by  
\ben
w(z) = w_0 + w_a \frac{z}{1+z},
\label{D1}
\een
where $w_0$ is the present value of the EoS and $w_a$ characterizes its redshift dependence. This is a basic extension of $w$CDM model~\cite{Suzuki, Planck5} where $w$ is dependent on $z$. 

We assume the total energy budget of the universe is composed of two non-interacting components: pressure less total matter with density $\rho_m$, and dark energy with density $\rho_d$ governed by the EoS in Eq.~(\ref{D1}).  
Under the assumption of a spatially flat FLRW background, the cosmic dynamics are determined by the Friedmann equations:
\ben
&&H^2 = \rho_m + \rho_d
\label{D2} \\
&&\frac{\Ddot{a}}{a} = -\frac{1}{6}\Big(\rho_m+ 3P_m + \rho_d + 3P_d \big)
\label{D3}
\een
The energy densities satisfy the continuity equation as follows:
\ben
&&\frac{d\rho_m}{dt} = -3 H\rho_m \n \\
&& \frac{d\rho_d}{dt} = -3 H \rho_d (1 + w).
\label{D4}
\een
We define ($\O_d$) and  ($\O_m$) as $\O_d= \frac{\rho_d}{3H^2}$, and $\O_m =\frac{\rho_m}{3 H^2}$ respectively. These lead to the differential equation for $H$ and $\O_m$ as,

\ben
&&\frac{dH}{dz} = \frac{3 H}{2(1+z)}(1 + (1 - \Omega_m) w)
\n \\
&& \frac{d\Omega_m}{dz} = -\frac{2 \Omega_m}{H}\frac{dH}{dz} + \frac{3\Omega_m}{1+z}
\label{D5}
\een
Together with Eq. (\ref{D5}), the condition $\Omega_{m0} + \Omega_{d0} =1$ specifies the total energy content of a spatially flat universe.

The set of Eqs.~(\ref{D5}),  (\ref{17}) along with the condition $\Omega_{m0} +\Omega_{d0} =1$, allows us to fully determine the evolution of the Hubble parameter and the energy content of the universe.

\begin{figure}[H] 
    \centering
    \includegraphics[width=\columnwidth]{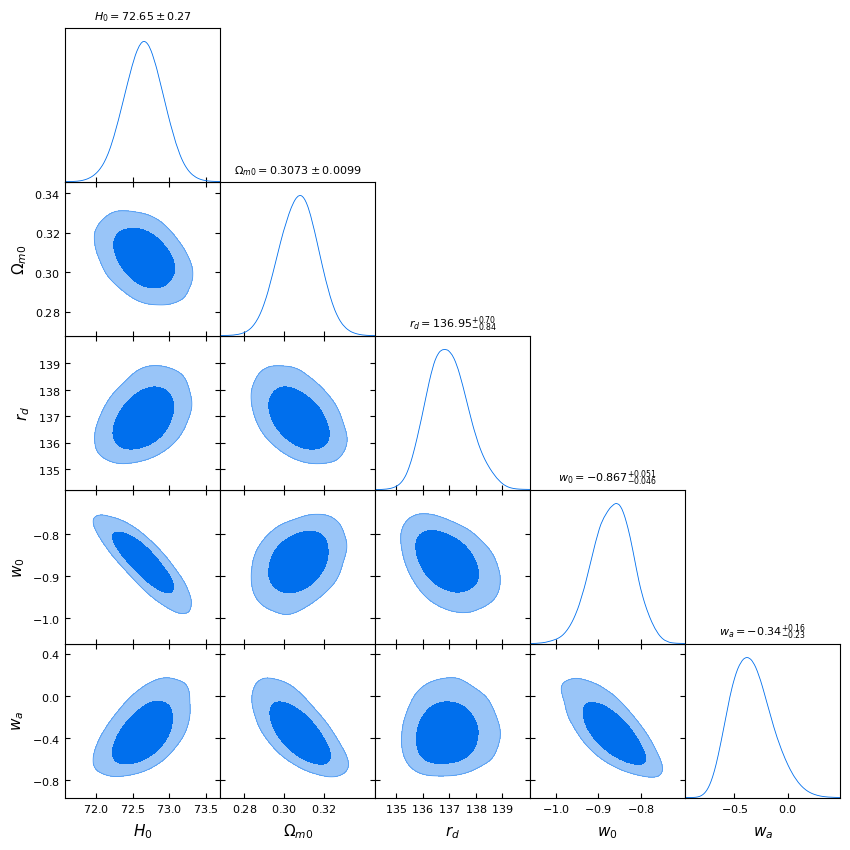}
    \caption{Bayesian analysis using SVI for the combined data set of Hubble+ Pantheon+SHOES BAO for $w$CDM Model}
    \label{Fig11} 
\end{figure}

\begin{table*}[ht]
\centering
\begin{tabular}{|c|c|c|}
\toprule
\textbf{Parameter} & \textbf{Prior} & \textbf{SVI best-fit} \\
\midrule
    $H_0$[Km/s/Mpc] & $\mathcal{U}(60,80)$ & $72.65 \pm 0.27$ \\
    \hline
    $\Omega_{m0}$ & $\mathcal{U}(0.1,0.5)$ & $0.3073 \pm 0.0099$ \\
    \hline
    $r_d$[Mpc] & $\mathcal{U}(100,200)$& $136.95^{+0.70}_{-0.84}$ \\
    \hline
    $w_0$ & $\mathcal{U}(-1.5, -0.5)$& $-0.867^{+0.051}_{-0.046}$ \\
    \hline
    $w_a$ & $\mathcal{U}(-1.0,1.0)$ & $-0.34^{+0.16}_{-0.23}$  \\
\bottomrule
\end{tabular}
\caption{Parameter estimates and uncertainties using Bayesian inference with NumPyro via SVI optimization for $w$CDM Model.}
\label{TableX}
\end{table*}

\begin{figure}[H] 
    \centering
    \includegraphics[width=\columnwidth]{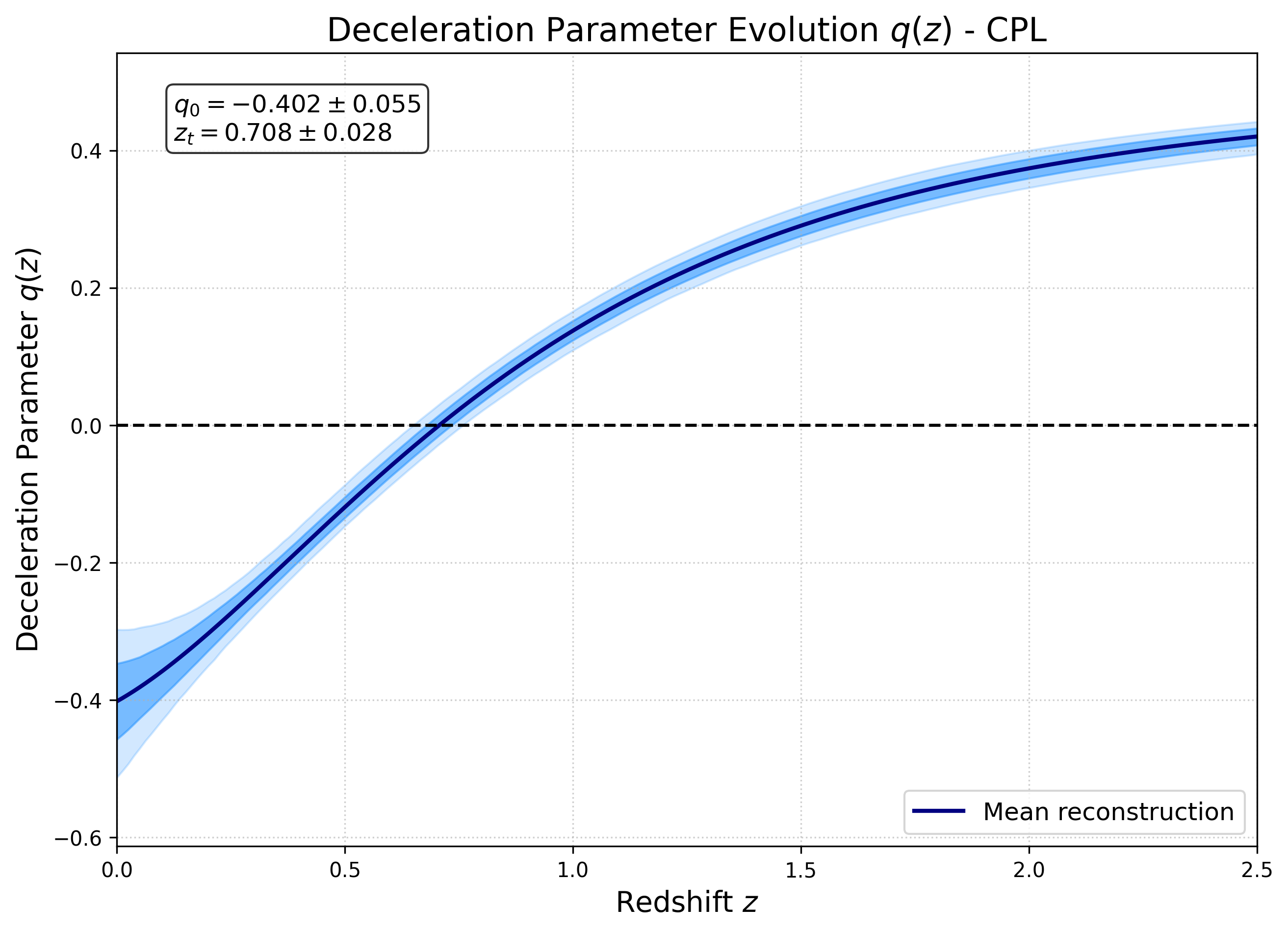}
    \caption{Plot of deceleration parameter~($q$) with redshift distance ($z$) for $w$CDM Model}
    \label{Fig12} 
\end{figure}

For the sake of completeness, we also plot the redshift dependence of the deceleration parameter, \(q(z)\), as a function of the cosmological redshift \(z\). Furthermore, we explicitly show the associated uncertainties in both the current value of the deceleration parameter, \(q_0\), and the transition redshift, \(z_t\). This plot and the associated error estimates are derived from the best fit cosmological parameters obtained via SVI based Bayesian inference. The numerical values of \(q_0\) and \(z_t\), together with their corresponding uncertainties, are reported in Fig.~\ref{Fig12}.

\section{Cosmological Parameter estimation of different Cosmological Model}
\label{E}

\begin{figure}[H] 
    \centering
   \includegraphics[width=\columnwidth]{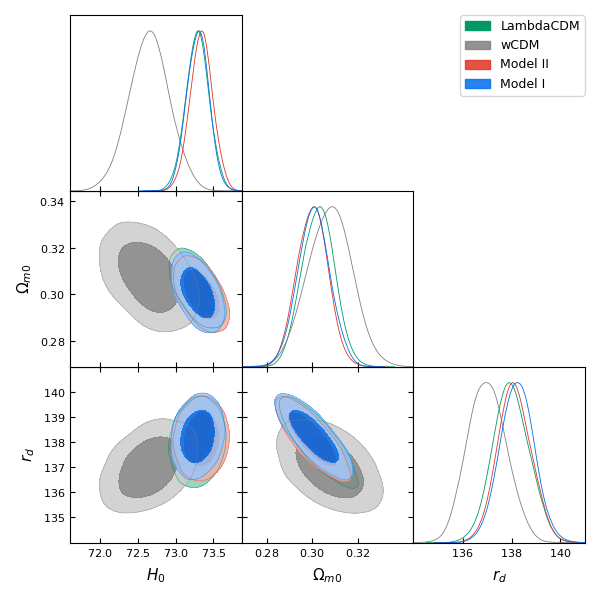}
    \caption{Bayesian analysis using SVI for the combined data set of Hubble+ Pantheon+SHOES BAO for all Models}
    \label{Fig13} 
\end{figure}

\begin{figure}[H] 
    \centering
    \includegraphics[width=\columnwidth]{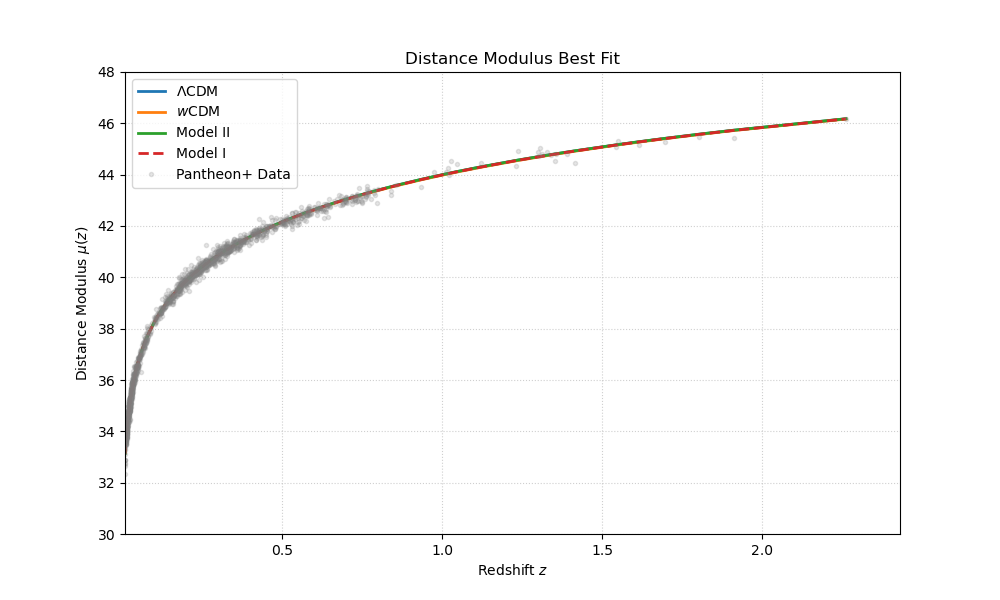}
    \caption{Comparison of Pantheon$+$SHOES data fits for all models.}
    \label{Fig15} 
\end{figure}

\begin{figure}[H] 
    \centering
    \includegraphics[width=\columnwidth]{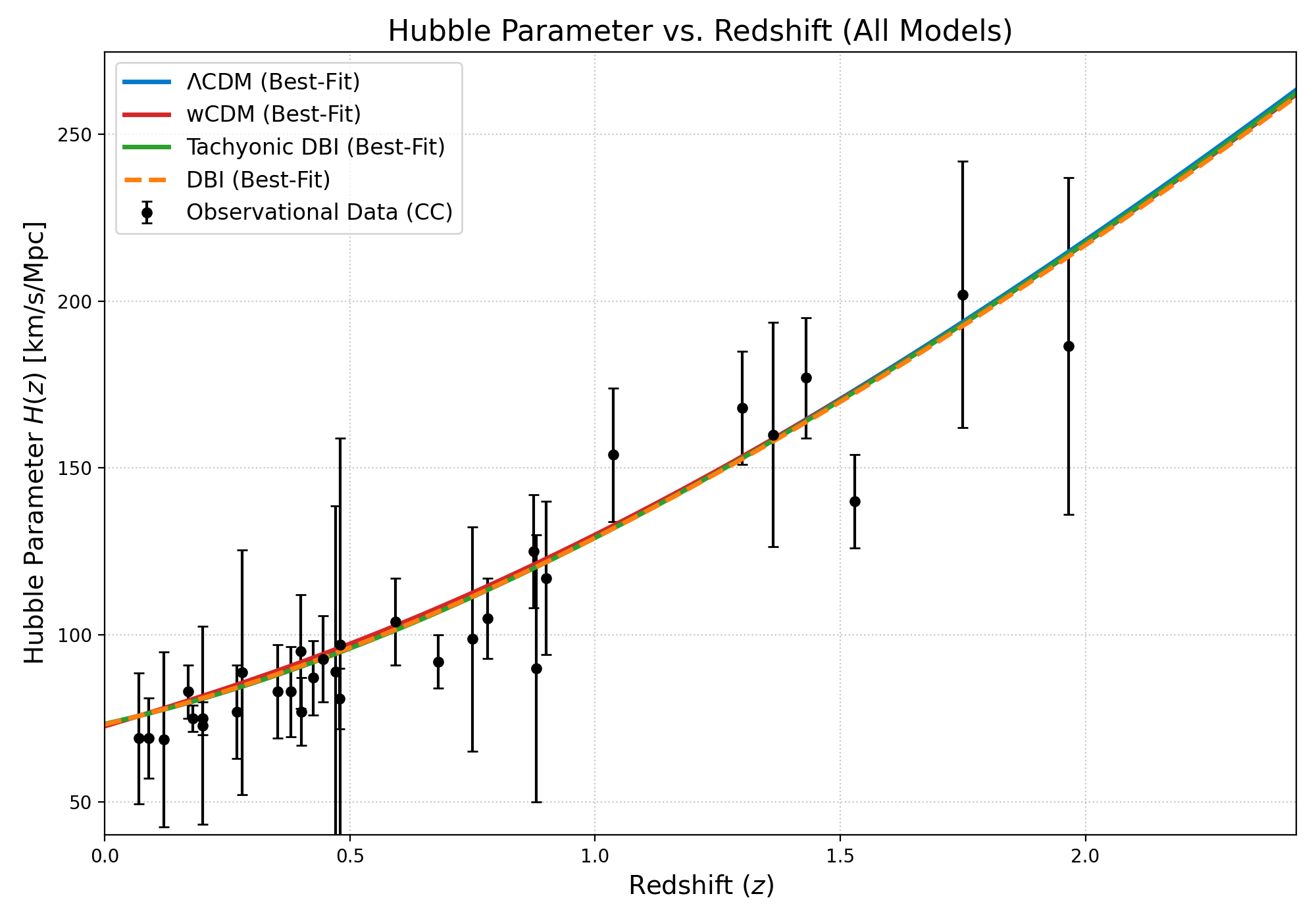}
    \caption{Comparison of Hubble cosmic chronometer data fits for all models.}
    \label{Fig14} 
\end{figure}

In this section, we plot SVI based bayesian analysis corner plot of all the cosmological parameter namely $H_0$, $\O_{m0}$ and $r_d$, which is provided in Fig.(\ref{Fig13}). We also assess the comparative performance of each model in fitting the Hubble cosmic chronometer data~(Fig.(\ref{Fig14})) and the combined Pantheon+SHOES supernova dataset~(Fig.(\ref{Fig15})).

\section{Dimensional Analysis of Two DBI Scalar Field Models}
\label{F}
We consider the DBI action for Model~I:
\ben
S_{DBI}^{I} &&= -\int d^4x \sqrt{-g} \,\Big[f(\phi)\sqrt{1- \frac{2X}{f(\phi)}} \n \\ &&- f(\phi) + V(\phi)\Big],
\label{F1}
\een
In natural units ($\hbar = 1$, $c = 1$), the action $S_{DBI}^{I}~([S]=0)$ is dimensionless, implying that the Lagrangian density has mass dimension $4$, i.e. $[\La]\equiv [M^4]$, with $[M] \equiv [E] \equiv [L^{-1}] \equiv [T^{-1}]$.
When $M_{pl}$ is treated as a dimension ($[M]$), the first Friedmann equation reads
\ben
3 M_{pl}^2 H^2 = \rho_{m} + \rho_{\phi} ,
\label{F2}
\een
where $\rho_{m}$ and $\rho_{\phi}$ denote the standard matter and scalar field energy densities, respectively.

Given that the left-hand side of Eq.~\eqref{F2} possesses a mass dimension of $[M^{4}]$, it follows that both the matter energy density $\rho_{m}$ and the scalar field energy density $\rho_{\phi}$ must similarly exhibit a dimension of $[M^{4}]$. 
Thus, based on Eqs.~\eqref{5} and \eqref{22}, dimensional consistency dictates that both the warp factor $f(\phi)$ and the potential $V(\phi)$ must possess a mass dimension of $[M^{4}]$.

Furthermore, the argument of the square root in Eq.~\eqref{F1} must be dimensionless, which implies
\ben
\left[\frac{2X}{f(\phi)}\right] = 1 .
\label{F3}
\een
Since $M_{pl}$ has mass dimension $[M]$ and $f(\phi)$ has dimension $[M^4]$, the kinetic term $X \equiv -\frac{1}{2} g^{\mu\nu}\partial_\mu\phi\,\partial_\nu\phi$ must carry dimension $[M^4]$. This requirement fixes the scalar field to have mass dimension one, $[\phi]=[M]$, as is standard for a dynamical scalar in four-dimensional relativistic quantum field theory.

Therefore, for the chosen forms \(f(\phi) = \lambda \phi^4\) and \(V(\phi) = \tfrac{1}{2} m^2 \phi^2\), the coupling \(\lambda\) is dimensionless, while \(m\) has mass dimension \([M]\).\\

For Model~II, the action for the scalar field is:
\ben
S_{DBI}^{II}= - \int d^4x \sqrt{-g} \bigg[ V(\phi)\sqrt{1- \frac{2X}{M_{pl}^4}}\bigg] .
\label{F4}
\een
Following the same argument as before, the action must be dimensionless in natural units $(\hbar=1,\;c=1)$, which implies that the corresponding Lagrangian density carries mass dimension $[M^4]$. Consequently, the argument of the square root must be dimensionless, requiring
\ben
\left[\frac{2X}{M_{pl}^4}\right]=1 .
\label{F5}
\een
This condition immediately implies that the kinetic term must satisfy $[X]=[M^4]$.

Recalling the definition of the kinetic term,
\ben
X \equiv -\frac{1}{2} g^{\mu\nu}\partial_{\mu}\phi\,\partial_{\nu}\phi ,
\label{F6}
\een
and noting that the metric tensor is dimensionless while spacetime derivatives carry mass dimension $[\partial_\mu]=[M]$, it follows that
\ben
[X]=[M^2][\phi]^2 .
\label{F7}
\een
Imposing the requirement $[X]=[M^4]$ uniquely fixes the scalar field to have mass dimension
\ben
[\phi]=[M] .
\label{F8}
\een

For a spatially homogeneous scalar field, the kinetic term reduces to
\ben
2X = \dot{\phi}^2 ,
\label{F9}
\een
which is fully consistent with the above dimensional assignment. Since the overall Lagrangian density must have mass dimension $[M^4]$, it follows that the potential $V(\phi)$ must also carry mass dimension $[M^4]$.

For the specific form of the potential adopted in our analysis,
\ben
V(\phi)=\frac{1}{2}m^2\phi^2 ,
\label{F10}
\een
dimensional consistency requires the mass parameter $m$ to carry mass dimension
\ben
[m]=[M] .
\label{F11}
\een
This is the standard result for a dynamical scalar field in four-dimensional relativistic quantum field theory.

More generally, in any local, Lorentz invariant relativistic field theory, a propagating scalar field in $D$ spacetime dimensions has mass dimension $[\phi]=(D-2)/2$, implying $[\phi]=1$ in four dimensions \cite{Schroeder,WeinbergQFT}. Non-canonical kinetic terms, such as those in k-essence or DBI theories, modify the dynamics but leave the canonical mass dimension of the field unchanged \cite{Picon1,Tong}.

However, in this study, we utilize Planck units by defining the reduced Planck mass $M_{\mathrm{pl}}=1$, thereby expressing all dimensionful quantities in terms of $M_{\mathrm{pl}}$. This selection simplifies the underlying equations and enhances numerical stability, while physical units can be easily restored by reintroducing the correct powers of $M_{\mathrm{pl}}$.

\end{document}